\documentclass[12pt,letterpaper]{article}
\usepackage[usenames, dvipsnames]{xcolor}
\usepackage{jcapmod}

\usepackage{tocloft}
\usepackage[]{todonotes}
\usepackage{verbatim}
\usepackage[utf8]{inputenc}
\usepackage{mathrsfs}
\usepackage{cleveref}
\usepackage[shortlabels]{enumitem}
\usepackage{setspace,caption}
\usetikzlibrary{external}
\makeatletter
\newsavebox\myboxA
\newsavebox\myboxB
\newlength\mylenA

\definecolor{cornellRed}{HTML}{B31B1B}
\definecolor{cornellBlue}{HTML}{0068AC}
\definecolor{cornellGreen}{HTML}{6EB43F}
\definecolor{light-gray}{gray}{0.95}

\usepackage{bbm} 					
\usepackage{slashed} 				
\usepackage{graphicx}				
\usepackage{subcaption}			
\usepackage{psfrag}				
\usepackage{tensor}				
\usepackage{fouridx}				
\usepackage{bm}					
\usepackage{mdframed}				
\usepackage{multirow}				
\usepackage{soul}					
\usepackage{bbold}				
\usepackage{multicol}				

\usepackage{amsmath}
\usepackage{amssymb}
\usepackage{amsfonts}
\usepackage{mathtools}
\usepackage[frozencache,cachedir=minted-cache-new]{minted} 

\usepackage{feynmf}
\usepackage{marvosym}

\usepackage{import}
\usepackage[linesnumbered,ruled,vlined]{algorithm2e}

\SetCommentSty{mycommfont}

\usepackage{multirow}
\usepackage{tabularx}
\usepackage{ragged2e}
\usepackage{booktabs}
\newcolumntype{Y}{>{\centering\arraybackslash}X}

\AtBeginDocument{
	\heavyrulewidth=.08em
	\lightrulewidth=.05em
	\cmidrulewidth=.03em
	\belowrulesep=.65ex
	\belowbottomsep=0pt
	\aboverulesep=.4ex
	\abovetopsep=0pt
	\cmidrulesep=\doublerulesep
	\cmidrulekern=.5em
	\defaultaddspace=.5em
}

\newcommand*\xoverline[2][0.75]{%
    \sbox{\myboxA}{$\m@th#2$}%
    \setbox\myboxB\null
    \ht\myboxB=\ht\myboxA%
    \dp\myboxB=\dp\myboxA%
    \wd\myboxB=#1\wd\myboxA
    \sbox\myboxB{$\m@th\overline{\copy\myboxB}$}
    \setlength\mylenA{\the\wd\myboxA}
    \addtolength\mylenA{-\the\wd\myboxB}%
    \ifdim\wd\myboxB<\wd\myboxA%
       \rlap{\hskip 0.5\mylenA\usebox\myboxB}{\usebox\myboxA}%
    \else
        \hskip -0.5\mylenA\rlap{\usebox\myboxA}{\hskip 0.5\mylenA\usebox\myboxB}%
    \fi}
\makeatother

\newcommand{\cT}{\mathcal{T}}

\definecolor{cobalt}{RGB}{44, 98, 120}
\definecolor{celadon}{rgb}{0.67, 0.88, 0.69}
\definecolor{dm}{cmyk}{.20, 0, .30, 0}
\definecolor{burgundy}{rgb}{0.5, 0.0, 0.13}
\definecolor{plotBlue}{RGB}{94, 130, 181}

\hypersetup{
  colorlinks,
  citecolor=Violet,
  linkcolor=cobalt,
  urlcolor=Blue}

\DeclareSymbolFontAlphabet{\mathbb}{AMSb}

\newif\iffastcompile

\fastcompilefalse

\iffastcompile
\newcommand{\cl}[1]{}
\newcommand{\lm}[1]{}
\newcommand{\md}[1]{}
\newcommand{\ab}[1]{}
\newcommand{\art}[1]{}
\else
\newcommand{\cl}[1]{\todo[color=burgundy!30, size=\scriptsize, bordercolor=burgundy!30]{CL: #1}}
\newcommand{\lm}[1]{\textcolor{red}{[LM: #1]}}
\newcommand{\md}[1]{\textcolor{blue}{[MD: #1]}}
\newcommand{\art}[1]{\textcolor{green}{[ART: #1]}}
\fi

\newcommand{\email}[1]{\href{mailto:#1}{#1}}

\ProvideTextCommandDefault{\Dbar}{%
\leavevmode\lower.5ex\rlap{\hskip-.07em\accent"16}D%
}

\begin{document}
	\newcommand{\main}{.}
\begin{titlepage}

\setcounter{page}{1} \baselineskip=15.5pt \thispagestyle{empty}
\setcounter{tocdepth}{2}

\bigskip\

\vspace{1cm}
\begin{center}

{{\fontsize{19}{26} \bfseries {\tt{CYTools:}} A Software Package for\\ \vskip 3.5pt
Analyzing Calabi-Yau Manifolds}}

\end{center}

\vspace{0.45cm}

\begin{center}
\scalebox{0.95}[0.95]{{\fontsize{14}{30}\selectfont  Mehmet Demirtas,${}^{a,b}$ Andres Rios-Tascon,${}^{c}$ and Liam McAllister${}^{c}$}}
 
\end{center}

\begin{center}

${}^a$ \textsl{The NSF AI Institute for Artificial Intelligence and Fundamental Interactions} \\
${}^b$ \textsl{Department of Physics, Northeastern University, Boston, MA 02115, USA} \\
${}^c$ \textsl{Department of Physics, Cornell University, Ithaca, NY 14853, USA}\\

\vspace{0.25cm}

\email{ m.demirtas@northeastern.edu, ar2285@cornell.edu,  mcallister@cornell.edu}
\end{center}

\vspace{0.6cm}
\noindent

We provide a user's guide to version 1.0 of the software package  {\tt{CYTools}}, which
we designed to compute the topological data of Calabi-Yau hypersurfaces in toric varieties.
{\tt{CYTools}} has strong capabilities in analyzing and triangulating polytopes, and can easily handle even the largest polytopes in the Kreuzer-Skarke list.
We explain the main functions and the options that can be used to optimize them, including example computations that illustrate efficient handling of large numbers of polytopes.
The software, installation instructions, and a Jupyter notebook tutorial can be found at \url{https://cy.tools}.

\noindent
\vspace{2.1cm}

\noindent\today

\end{titlepage}
\tableofcontents\newpage

\section{Introduction}

Calabi-Yau compactifications provide a unique proving ground for ideas about quantum gravity and geometry.
Although supersymmetry may have little to do with the low-energy physics of our universe, it provides unparalleled theoretical control, and brings within reach a host of otherwise-inaccessible quantum gravity questions.  Through careful study of compactifications of critical superstrings, M-theory, and F-theory on Calabi-Yau $n$-folds, one can learn a great deal about the physics of the Planck scale.

Working even with simple Calabi-Yau threefolds requires efforts in algebraic geometry that can be a stretch for many physicists, presenting a barrier to entry into the subject.  Moreover, most known threefolds have such topological complexity that some of their properties are out of reach for anyone equipped only with pencil and paper --- indeed, until recently these properties were also inaccessible to computers.

Nevertheless, an astronomically large set of Calabi-Yau threefolds can \emph{in principle} be analyzed through purely mechanical computation from a combinatorial starting point.
Batyrev \cite{Batyrev:1994hm} has shown how to construct Calabi-Yau threefold hypersurfaces in toric varieties in terms of triangulations of 4-dimensional reflexive polytopes, and Kreuzer and Skarke \cite{Kreuzer:2000xy} have classified and listed all such polytopes.  To manufacture and analyze threefolds, one needs only to follow Batyrev's prescription applied to the Kreuzer-Skarke list.

The difficulty is a practical one: most triangulations of polytopes in the Kreuzer-Skarke list, and thus most candidate Calabi-Yau threefold hypersurfaces, are inaccessible by existing publicly-available codes, even on contemporary hardware.
As we will explain, the problem is not that there is a single, fundamental computation of irreducible complexity.  Instead, many interrelated computations each become extremely slow for polytopes with many points, corresponding to Calabi-Yau threefolds in which the number of moduli, given by the Hodge numbers $h^{1,1}$ and $h^{2,1}$, is large.
General purpose algorithms that are not crafted specifically for dealing with Calabi-Yau hypersurfaces in toric varieties are typically effective only when the relevant Hodge number --- for example, $h^{1,1}$ for a computation of triple intersection numbers of divisors, or $h^{2,1}$ for a computation of periods of the holomorphic $(3,0)$ form --- is $\lesssim 10$.

The purpose of the software package {\tt{CYTools}} is to overcome this limitation, and enable efficient study of an immense class of Calabi-Yau manifolds.
To achieve this, we examined the series of computational steps connecting a polytope to a Calabi-Yau threefold.  We optimized existing algorithms, or developed new algorithms, until each step took at most a few milliseconds for polytopes of modest size, and at most of order one second for the largest polytopes in the Kreuzer-Skarke list.

\texttt{CYTools} primarily consists of a Python package that contains a variety of algorithms and tools designed to analyze Calabi-Yau hypersurfaces in toric varieties. It makes use of some open-source software to aid with particular computations that are already well-optimized or would be too slow to run with Python. We have written Python classes that interface with the external software, and handle all the different steps in the computation, starting from polytope information, then obtaining suitable triangulations, and finally computing various properties of the resulting toric varieties and Calabi-Yau hypersurfaces. All the functionality can be easily accessed with intuitive Python functions, and it can be used with Jupyter notebooks, allowing beginner and expert programmers to suit their needs.

The organization of this user's guide is as follows.
In \S\ref{sec:overview} we explain how to install \texttt{CYTools}, and then briefly review key concepts and terminology for Calabi-Yau hypersurfaces in toric varieties.
In \S\ref{sec:coreobjects} we demonstrate how \texttt{CYTools} operates on the data of polytopes, triangulations, and Calabi-Yau hypersurfaces.
In \S\ref{sec:examples} we give a step-by-step walkthrough of a few examples.
In \S\ref{sec:methods} we
describe some of the key algorithms used in \texttt{CYTools}.
We discuss potential applications, and directions for future development of \texttt{CYTools}, in \S\ref{sec:outlook}.

The \texttt{CYTools} software can be downloaded from
\url{https://cy.tools}.
This website also contains the most detailed and up-to-date installation instructions, troubleshooting steps, instructions for advanced usage, and a Jupyter notebook with a walkthrough tutorial.

\begin{figure}[!ht]
    \centering
    \includegraphics[width=5cm]{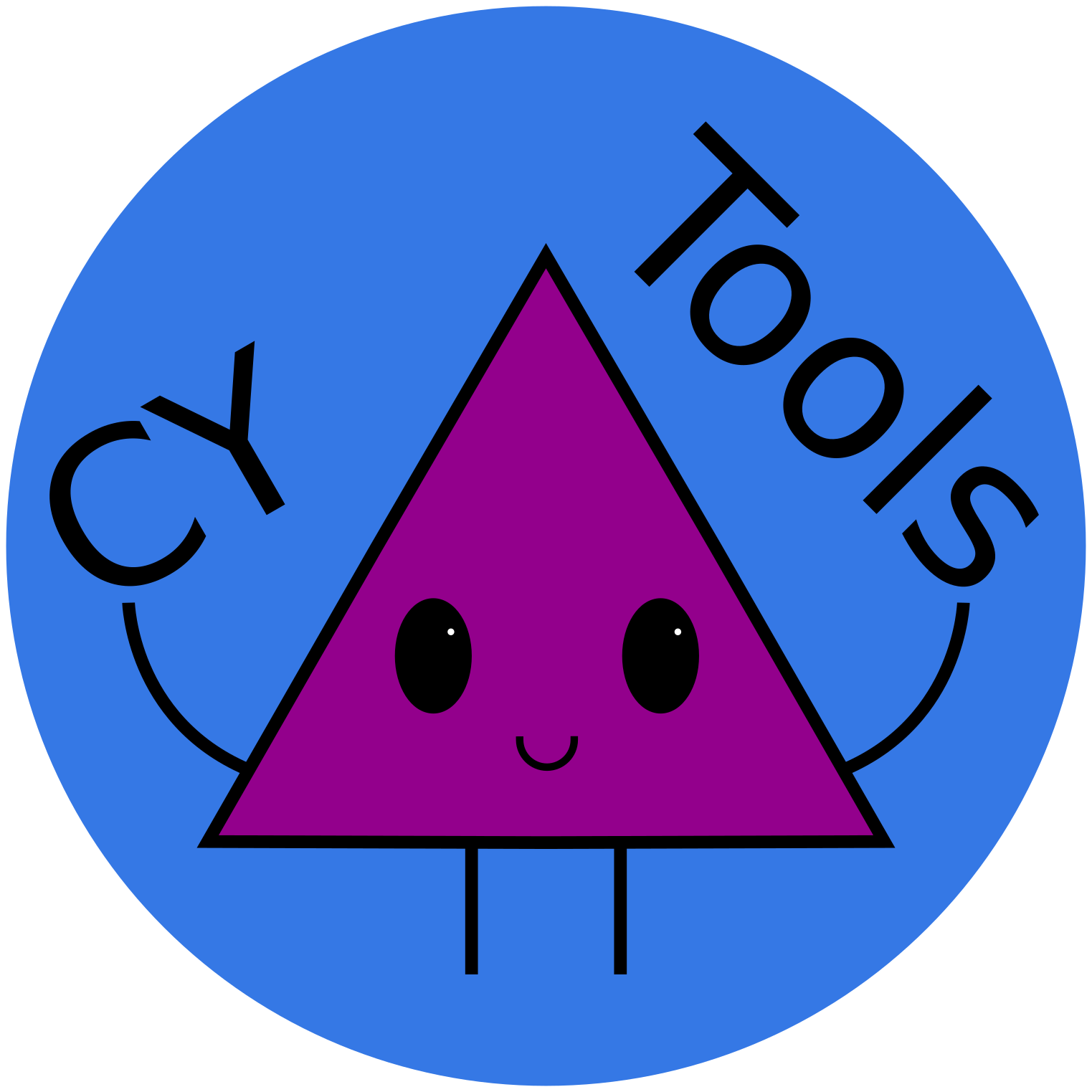}
	\label{fig:cytools_logo}
\end{figure}

\section{Overview} \label{sec:overview}
In this section we provide a beginner's guide to installing and using \texttt{CYTools}, and then review how Calabi-Yau hypersurfaces are constructed from triangulations of reflexive polytopes.

\subsection{Getting started}

The installation of \texttt{CYTools} is done in the form of a Docker image. Docker \cite{merkel2014docker} is a tool that allows software to be packaged into an image that contains all the dependencies and configuration necessary to run it. These images are used to build containers, which are essentially lightweight, self-contained virtual machines
that allow the software to run in an environment isolated from the host operating system. This allows \texttt{CYTools} to be easily installed across all major operating systems.

On a computer with an existing Docker installation,\footnote{See \url{https://docs.docker.com/} for Docker installation instructions.} installing \texttt{CYTools} takes either a single line in the terminal, or the execution of an installer script. On Linux and macOS it can be installed by running the following line in the terminal.
\begin{minted}[
bgcolor=light-gray,
fontsize=\footnotesize
]{bash}
$ curl https://cy.tools/install.sh | bash
\end{minted}
On Windows computers, it can be installed by running the installer script found at \url{https://cy.tools/cytools-installer.bat}.

Once installed, launching \texttt{CYTools} simply requires double-clicking on the icon that should appear on your start menu (or equivalent) after installation. Alternatively, it can be launched from the terminal with the command \mintinline{bash}{cytools}. In either case, a splash screen and a handful of logs will briefly flash, and then a new tab will be opened in the default web browser. In rare cases when the browser does not open, there will be a link shown that you can either ctrl+click, or copy and paste into your browser of choice. You will be greeted with a standard JupyterLab environment where you can edit files, launch terminals, and, most importantly, work with Jupyter notebooks. If you are not familiar with JupyterLab or Jupyter notebooks, you can find plenty of resources online.\footnote{See e.g. \url{https://jupyterlab.readthedocs.io/en/stable/}.} For the rest of our discussion we will assume basic familiarity with JupyterLab and Python.

Docker usage generally requires administrative privileges on the host machine. Since this is not always possible, for example when using a shared cluster at a university, we have also made it possible to install \texttt{CYTools} using Singularity. Singularity \cite{10.1371/journal.pone.0177459} is an alternative to Docker that is more focused on scientific computing, and that is commonly installed on computing clusters since its use does not require administrative privileges. However, it is only available on Linux, it is more difficult to install, and there are fewer online resources for it. For these reasons, Docker is the primary installation method.
Advanced users seeking to modify the Docker image or to use \texttt{CYTools} on a cluster with Singularity can find guidance at
\url{https://cy.tools}.

\subsection{Calabi-Yau manifolds from reflexive polytopes}
We now give a brief summary of Batyrev's construction of Calabi-Yau hypersurfaces in toric varieties. Our goal is to provide a practical overview  and set notation.
For a more in-depth exposition, we refer the reader to \cite{Batyrev:1994hm,Greene:1996cy,cox2011toric}.
A limited glossary of terms and definitions appears in Appendix \ref{appx:definitions}. We will focus on Calabi-Yau threefolds for concreteness, but analogous statements hold for Calabi-Yau fourfolds, and \texttt{CYTools} is capable of performing the
corresponding computations.

Let $M$, $N$ be a pair of dual lattices isomorphic to $\mathbb{Z}^4$, let $\Delta \in M$ be a 4-dimensional reflexive lattice polytope, and let $\Delta^\circ \in N$ be its dual. The normal fan of $\Delta$, or equivalently the face fan of $\Delta^\circ$, defines a toric variety with a possibly singular anticanonical hypersurface. A smooth anticanonical hypersurface can be obtained by finding a suitable refinement of the fan that corresponds to a maximal projective crepant partial desingularization. This amounts to obtaining a  regular, star, and sufficiently fine triangulation of the lattice points on $\Delta^\circ$. Specifically, points strictly interior to facets correspond to divisors in the toric variety that do not intersect a generic anticanonical hypersurface, and can safely be ignored.  So suppose that $\mathscr{T}$ is a regular, star triangulation of $\Delta^\circ$ such that every point of $\Delta^\circ$ that is not strictly interior to a facet is a vertex of a simplex in $\mathscr{T}$.  Then, although strictly speaking one should call
$\mathscr{T}$ `fine except for points interior to facets', we will abuse terminology and say that $\mathscr{T}$ is \emph{fine}, and is an FRST.
The toric variety arising from such an FRST has a smooth anticanonical hypersurface that is Calabi-Yau \cite{Batyrev:1994hm}.

Famously, the full list of 473,800,776 4-dimensional reflexive polytopes was constructed by Kreuzer and Skarke (KS) \cite{Kreuzer:2000xy}, and is available in a database that now bears their name \cite{KSdatabase}. While 473,800,776 is not a small number, {\tt{CYTools}} can easily process all polytopes in the  Kreuzer-Skarke database. The number of FRSTs per polytope, however, increases \textit{exponentially} with the number of lattice points on the polytope. The number of topologically distinct Calabi-Yau threefold hypersurfaces is not known, but it could be as high as $10^{428}$ \cite{Demirtas:2020dbm}. Enumerating all Calabi-Yau hypersurfaces is therefore infeasible.
Even so, {\tt{CYTools}}
enables the construction and analysis of large and diverse samples\footnote{See \cite{Demirtas:2020dbm} and \S\ref{sss:sample} for an explanation of sampling algorithms.} of Calabi-Yau hypersurfaces from the entire database of polytopes.

Once a Calabi-Yau hypersurface is defined by specifying a polytope and an FRST, the next step is to compute geometric quantities of interest. Among the most important topological invariants are the Hodge numbers $h^{1,1}$ and $h^{2,1}$, and the Euler characteristic $\chi = 2(h^{1,1} - h^{2,1})$. These depend only on polytope data:

\begin{eqnarray}
h^{1,1}(X) &=& \sum_{\Theta^\circ \in {\cal F}(\le 2)}\ell^*(\Theta^\circ) - 4  + \sum_{\Theta^\circ \in {\cal F}(2)} \ell^*(\Theta^\circ)\ell^*(\Theta)\,,\label{eq:h11}\\
h^{2,1}(X) &=& \,\sum_{\Theta \in {\cal F}(\le 2)}\ell^*(\Theta) ~\,- 4  + \,\sum_{\Theta \in {\cal F}(2)} \ell^*(\Theta^\circ)\ell^*(\Theta)\,,\label{eq:h21}
\end{eqnarray}
where $\Theta$ denotes a face of $\Delta$ and $\Theta^\circ$ its dual; $\mathcal{F}(d)$ and $\mathcal{F}(\leq d)$ denote the set of all faces with dimension $d$ and $\leq d$, respectively; and $\ell^*(\Theta)$ is the number of points strictly interior to the face $\Theta$. Polytopes for which the final term in \eqref{eq:h11} vanishes are called \emph{favorable}.  Checking favorability is immediate in \texttt{CYTools}, as we will demonstrate in \S\ref{sec:examples}.
Much of the functionality of \texttt{CYTools} is automatic for favorable polytopes but requires a certain amount of adaptation for non-favorable cases.  We intend to make the treatment of non-favorable polytopes more user-friendly in a future release.

Each ray in the toric fan (equivalently, each lattice point on the boundary of $\Delta^\circ$) corresponds to a prime toric divisor $\hat{D}^a$ of the toric variety. The degrees in the Cox ring, more commonly referred to as the GLSM charge matrix in the physics literature, encode the linear relationships among these divisors. As mentioned before, toric divisors that correspond to lattice points not strictly interior to facets intersect a generic Calabi-Yau hypersurface $X$, giving rise to a divisor in $X$, $D^a = \hat{D}^a \cap X$. The divisors $D^a$, of which there are $h^{1,1}+4$ in the case of a favorable polytope,
generate $H_4(X, \mathbb{Z})$. We will often pick $h^{1,1}$ of these divisors (selecting a set that are linearly independent over $\mathbb{R}$) to form a basis of $H_4(X, \mathbb{Z})$.\footnote{Every such choice will provide a basis over $\mathbb{Q}$, and some also furnish a basis over $\mathbb{Z}$.  In practice we have always been able to find a choice that furnishes a $\mathbb{Z}$-basis, but we are not aware of a proof that such a basis always exists.}

We denote the intersection numbers of the $D^a$ as $\kappa_{abc}$,
\begin{equation}
    \kappa_{abc} = \# D^a \cap D^b \cap D^c .
\end{equation}
The intersection numbers are important data for string compactifications, in part because they define a product in the de Rham cohomology of the Calabi-Yau manifold, and computing the volumes of cycles calibrated by the K\"ahler form requires computing the intersection numbers. One of the most important individual advances provided by \texttt{CYTools} is the capability to compute intersection numbers in at most a second even for Calabi-Yau hypersurfaces with many moduli.

The homology classes in $H_2(X,\mathbb{Z})$ and $H_4(X,\mathbb{Z})$ that have holomorphic representatives are called effective. The sets of effective classes in $H_2(X,\mathbb{Z})$ and $H_4(X,\mathbb{Z})$ are called the Mori cone (or cone of effective curves) and the effective cone (or cone of effective divisors), respectively.
A holomorphic representative of an effective homology class is volume-minimizing in its class, and its volume can be obtained by integrating the appropriate power of the K\"ahler form over the cycle.
As the action of an extended object wrapping a cycle is proportional to the cycle volume to a good approximation in the geometric regime, identifying volume-minimizing cycles is an important problem. Additionally, the structure of the superpotential in superstring compactifications depends strongly on the holomorphic cycles in the compactification. The Mori cone of a Calabi-Yau manifold is of special interest, as its dual is the (closure of the) K\"ahler cone, the space of K\"ahler forms.

There is no general algorithm known to us that allows for computing the cones of effective cycles of Calabi-Yau hypersurfaces with large Hodge numbers.\footnote{While it is sometimes possible to determine whether a given homology class is effective, such computations become infeasible when $h^{1,1} \gg 10$.} Fortunately, analogous computations for the ambient toric variety are conceptually straightforward, and provide useful approximations to the Mori and effective cones of the Calabi-Yau hypersurface: see \cite{Demirtas:2018akl}.

While we have limited our discussion to even-dimensional cycles and the K\"ahler moduli space of Calabi-Yau manifolds so far, periods of 3-cycles and the complex structure moduli space are characterized by essentially the same data described above, via mirror symmetry. Utilizing mirror symmetry to compute periods requires additional computational methods, and will be discussed in future work \cite{computational-mirror-symmetry}.

\section{Calabi-Yau Hypersurfaces in \texttt{CYTools}}\label{sec:coreobjects}

Let us walk through the process of analyzing a Calabi-Yau manifold in \texttt{CYTools}, starting from a polytope, finding a triangulation, and then constructing the corresponding toric variety and Calabi-Yau hypersurface. Most of the code blocks in this section can also be found in the tutorial page in the documentation website, where we focus on code usage as opposed to explaining the underlying mathematical objects.

\subsection{Polytopes}

Batyrev's construction of Calabi-Yau manifolds begins with reflexive polytopes. Polytope objects can be constructed in \texttt{CYTools} with the \mintinline{python}{Polytope} class. Only lattice polytopes are supported. As a consequence, any floating-point numbers that are used to construct a polytope are truncated to integers.

Let us look at a simple example where we import the class, construct a square, and print basic information about it.

\begin{minted}[
%frame=lines,
%framesep=2mm,
%baselinestretch=1.2,
bgcolor=light-gray,
fontsize=\footnotesize,
%linenos,
escapeinside=||
]{python}
>>> from cytools import Polytope
>>> p = Polytope([[1,0],[0,1],[-1,0],[0,-1]])
>>> print(p)
|\text{A 2-dimensional reflexive lattice polytope in ZZ\textsuperscript{$\wedge$}2}|
\end{minted}

The \mintinline{python}{Polytope} class contains numerous functions to perform all the main computations one typically needs. The functions have self-explanatory names whenever possible. For example, we can compute the dimension of the polytope, and whether it is reflexive, as follows.
\begin{minted}[
bgcolor=light-gray,
fontsize=\footnotesize,
escapeinside=||
]{python}
>>> p.dimension()
|2|
>>> p.is_reflexive()
|True|
\end{minted}
The lattice points of the polytope can be found as follows.
\begin{minted}[
bgcolor=light-gray,
fontsize=\footnotesize,
escapeinside=||
]{python}
>>> p.points()
|array([[ 0,  0],
       [-1,  0],
       [ 0, -1],
       [ 0,  1],
       [ 1,  0]])|
\end{minted}
As a last example for now, here is how one can find the 1-faces of the polytope.
\begin{minted}[
bgcolor=light-gray,
fontsize=\footnotesize,
escapeinside=||
]{python}
>>> p.faces(d=1)
|(A 1-dimensional face of a 2-dimensional polytope in ZZ\textsuperscript{$\wedge$}2,
 A 1-dimensional face of a 2-dimensional polytope in ZZ\textsuperscript{$\wedge$}2,
 A 1-dimensional face of a 2-dimensional polytope in ZZ\textsuperscript{$\wedge$}2,
 A 1-dimensional face of a 2-dimensional polytope in ZZ\textsuperscript{$\wedge$}2)|
\end{minted}
This function returns a tuple of \mintinline{python}{PolytopeFace} objects, with functions similar to those of the \mintinline{python}{Polytope} class. As we will see, this pattern of functions returning more refined objects is ubiquitous in \texttt{CYTools}. This makes the process of constructing Calabi-Yau manifolds more intuitive and reliable, as the functions make sure that all the data uses the right conventions for the computations.

\texttt{CYTools} uses efficient algorithms that substantially improve upon previous open source software packages in the literature, especially in handling 4-dimensional reflexive polytopes.\footnote{For some computations involving 5-dimensional polytopes we have not improved on the performance of \texttt{PALP} \cite{Kreuzer:2002uu}, but we note that \texttt{CYTools} provides a convenient interface to \texttt{PALP}.} In particular, we have drastically increased the performance for computing the lattice points and faces of the polytopes. We will describe these algorithms in detail in  \S\ref{sec:lattice-points} and \S\ref{sec:faces}.

\subsubsection{Kreuzer-Skarke database}

As mentioned above, 4-dimensional reflexive polytopes are the starting point for obtaining Calabi-Yau threefolds using Batyrev's construction.
\texttt{CYTools} contains the \mintinline{python}{fetch_polytopes} function, which provides a convenient way to fetch polytopes from the Kreuzer-Skarke  database. The number of points, vertices, or facets of the polytope, or the Hodge numbers of the resulting Calabi-Yau hypersurfaces, among other parameters, can be specified. As an example, we obtain 100 polytopes with 5 vertices that produce Calabi-Yau hypersurfaces with $h^{1,1}=27$ when triangulated (i.e.~the polytopes live in the $N$ lattice).
\begin{minted}[
bgcolor=light-gray,
fontsize=\footnotesize,
escapeinside=||
]{python}
>>> from cytools import fetch_polytopes
>>> gen = fetch_polytopes(n_vertices=5, h11=27, lattice="N", limit=100)
>>> print(gen) # The number below will vary, as it is a memory address
|<generator object polytope\_generator at 0x7f306eacaeb0>|
\end{minted}
Note that this function returns a generator object instead of a list. This is done for performance reasons, as is explained in more detail on the website. We also mention that when using Hodge numbers of a polytope, as in the above function, we must specify whether we want to interpret the polytope as being in the $M$ lattice, i.e.~as being the Newton polytope of the Calabi-Yau hypersurface, or as being in the $N$ lattice, i.e.~as being the polytope for which a suitable refinement of its face fan defines the ambient toric variety. We enforce this requirement because in parts of the literature it is common to interpret the polytope in the $M$ lattice, while in others it is more common to interpret it in the $N$ lattice, so it is a frequent source of confusion. A helpful mnemonic \cite{Bouchard:2007ik} is that points on the $N$-lattice polytope are the rays of the toric fa\textbf{n} corresponding to divisors, while points in the $M$-lattice polytope correspond to
\textbf{m}onomials in the defining polynomial of the Calabi-Yau hypersurface. Thus, a polytope $\mathcal{P}$ with many points gives rise to a Calabi-Yau threefold $X$ with many prime toric divisors, and hence many K\"ahler moduli, if $\mathcal{P}$ is taken to live in the $N$ lattice.  If instead $\mathcal{P}$ is taken to live in the $M$ lattice, one finds the mirror threefold $\tilde{X}$, which has many monomials in its defining equation, and hence many complex structure moduli.

\texttt{CYTools} also includes the \mintinline{python}{read_polytopes} function, which reads from a string or input file containing polytopes in the format used by the Kreuzer-Skarke  database. The function \mintinline{python}{read_polytopes} is similar to \mintinline{python}{fetch_polytopes}, but is recommended when performing large scans, as \mintinline{python}{fetch_polytopes} may be quite slow when obtaining a large number ($\gtrsim 10,000$) of polytopes, especially when being very specific about their properties.

We provide further examples of these functions in \S\ref{sec:examples}, and the full list of parameters can be found in the documentation website \cite{cytools-website}.

\subsection{Triangulations}\label{sec:triangulations}

Triangulations of polytopes are handled in \texttt{CYTools} by the \mintinline{python}{Triangulation} class, which contains a variety of functions to compute many of the relevant properties of triangulations, and also serves to interface with external software that performs other computations. Let us see how we can obtain triangulations of polytopes and analyze their properties.

\subsubsection{Obtaining a triangulation}\label{sec:obtaining_triangulation}

A triangulation of a lattice polytope can be obtained using the \mintinline{python}{triangulate} function of the \mintinline{python}{Polytope} class. As a first example, we can construct a polytope and obtain a triangulation of it as follows.

\begin{minted}[
bgcolor=light-gray,
fontsize=\footnotesize,
escapeinside=||
]{python}
>>> p = Polytope([[1,0,0,0],[0,1,0,0],[0,0,1,0],[0,0,0,1],[-1,-1,-6,-9]])
>>> t0 = p.triangulate()
>>> print(t0)
|A fine, regular, star triangulation of a 4-dimensional point configuration with
7 points in ZZ\textsuperscript{$\wedge$}4|
\end{minted}
When the polytope is reflexive, the \mintinline{python}{triangulate} function by default constructs a fine, regular, star triangulation of \emph{the points not strictly interior to facets}.
The set of points to be included in the triangulation can be specified explicitly by modifying the input parameters of \mintinline{python}{triangulate}, as we will describe below.

The \mintinline{python}{triangulate} function returns a \mintinline{python}{Triangulation} object with useful functions of its own. For example, there are functions to check if it is fine (with respect to a given point configuration, that by default does not include points interior to facets), regular, or star.
\begin{minted}[
bgcolor=light-gray,
fontsize=\footnotesize,
escapeinside=||
]{python}
>>> t.is_fine()
|True|
>>> t.is_regular()
|True|
>>> t.is_star()
|True|
\end{minted}
We demonstrate a number of other useful functions in \S\ref{sec:examples}, and as usual refer the reader to the documentation website \cite{cytools-website} for the full list of functions.

\begin{figure}[!ht]
    \centering
    \includegraphics[width=\textwidth]{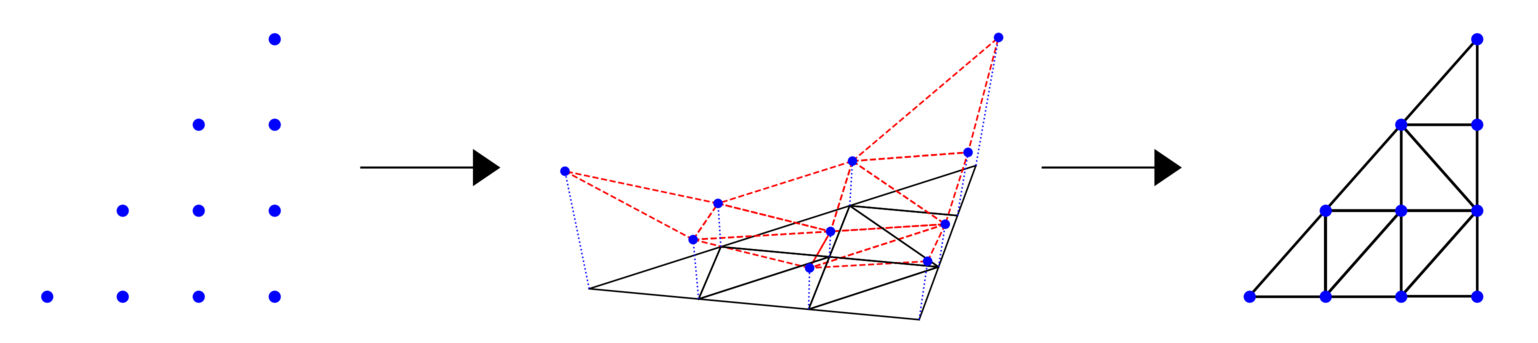}
	\caption{Regular triangulations descend from projections of the lower faces of a convex hull in one higher dimension. In the figure, a 2-dimensional point set is lifted by a set of heights into a 3-dimensional convex hull. The lower faces (shown in red) induce a triangulation on the original point set via their projection.  Figure from \cite{Demirtas:2020dbm}.}
	\label{fig:reg_triang}
\end{figure}

\texttt{CYTools} utilizes multiple external software packages to obtain triangulations. The main method it uses is to lift the points into an extra dimension by a set of heights, compute the convex hull of these points, and then project down the lower faces of the resulting polyhedron. The process is illustrated in Fig.~\ref{fig:reg_triang}. This method is considerably faster than the pushing algorithm used in \texttt{TOPCOM} \cite{Rambau2002}, which is partly why \texttt{CYTools} can vastly outperform standard mathematical software like \texttt{SageMath} \cite{sagemath}. However, although all regular triangulations can be obtained from this procedure, it does not allow a systematic enumeration of them. Thus, \texttt{CYTools} still relies on \texttt{TOPCOM} for some computations. We will discuss the algorithms in more detail in \S\ref{sec:methods_triangulations}.

The \mintinline{python}{triangulate} function has various parameters that can be included as input. For example, one can force the inclusion of points interior to facets, one can input a set of heights, or even input the simplices of a pre-computed triangulation. One can include points interior to facets as follows.
\begin{minted}[
bgcolor=light-gray,
fontsize=\footnotesize,
escapeinside=||
]{python}
>>> p = Polytope([[1,0,0,0],[0,1,0,0],[0,0,1,0],[0,0,0,1],[-1,-1,-6,-9]])
>>> t1 = p.triangulate(include_points_interior_to_facets=True)
>>> print(t1)
|A fine, regular, star triangulation of a 4-dimensional point configuration with
10 points in ZZ\textsuperscript{$\wedge$}4|
\end{minted}
Note that now  the point configuration contains 10 lattice points instead of the 7 that were included in the triangulation \mintinline{python}{t0} that we constructed above. If we want to construct a triangulation with a given set of heights we can do so as follows. We also show how to force the resulting triangulation to be star.
\begin{minted}[
bgcolor=light-gray,
fontsize=\footnotesize,
escapeinside=||
]{python}
>>> p = Polytope([[1,0,0,0],[0,1,0,0],[0,0,1,0],[0,0,0,1],[-1,-1,-6,-9]])
>>> t2 = p.triangulate(heights=[10,2,3,4,5,6,7])
>>> print(t2)
|A non-fine, regular, non-star triangulation of a 4-dimensional point configuration
with 7 points in ZZ\textsuperscript{$\wedge$}4|
>>> t3 = p.triangulate(heights=[10,2,3,4,5,6,7], make_star=True)
>>> print(t3)
|A non-fine, regular, star triangulation of a 4-dimensional point configuration
with 7 points in ZZ\textsuperscript{$\wedge$}4|
\end{minted}
Finally, we show how to construct a triangulation from a list of simplices that was previously found. In fact, it is the same triangulation as \mintinline{python}{t3}.
\begin{minted}[
bgcolor=light-gray,
fontsize=\footnotesize,
escapeinside=||
]{python}
>>> p = Polytope([[1,0,0,0],[0,1,0,0],[0,0,1,0],[0,0,0,1],[-1,-1,-6,-9]])
>>> t4 = p.triangulate(simplices=[[0,1,2,3,4],[0,1,2,3,5],[0,1,2,4,5],
>>>                               [0,1,3,4,5],[0,2,3,4,5]])
>>> print(t4)
|A non-fine, regular, star triangulation of a 4-dimensional point configuration
with 7 points in ZZ\textsuperscript{$\wedge$}4|
\end{minted}

\subsubsection{Obtaining all triangulations}

For polytopes with a small number of lattice points, it is possible to compute the full set of triangulations. \texttt{CYTools} achieves this with the use of \texttt{TOPCOM}. We provide a very simple interface with this external software: finding all triangulations only requires a single line.
\begin{minted}[
bgcolor=light-gray,
fontsize=\footnotesize,
escapeinside=||
]{python}
>>> p = Polytope([[1,0,0,0],[0,1,0,0],[0,0,1,0],[0,0,0,1],[-1,-1,-6,-9]])
>>> gen = p.all_triangulations()
>>> t = next(gen)
>>> print(t)
|A fine, regular, star triangulation of a 4-dimensional point configuration with
7 points in ZZ\textsuperscript{$\wedge$}4|
\end{minted}
This function takes various arguments to indicate whether the triangulations should be fine, regular, and/or star. As with \mintinline{python}{fetch_polytopes}, the \mintinline{python}{all_triangulations} function returns a generator object for performance reasons. The triangulations can then be obtained with the \mintinline{python}{next} function as above, or with a for-loop, as will be shown later.

We have now seen how to construct a single triangulation, or all triangulations of a polytope. However, for most of the polytopes in the Kreuzer-Skarke database the number of triangulations is exponentially large, so listing all of their triangulations is impossible. Nevertheless, one certainly would want to obtain more than a single triangulation, so we now explain how to construct large sets of suitably random triangulations.

\subsubsection{Obtaining random triangulations}

Obtaining ensembles of triangulations is important for a variety of applications, ranging from training neural networks to compute topological properties of the Calabi-Yau very quickly \cite{Demirtas:2020dbm}, to studying distributions of observables in the string landscape. Particularly for the latter purpose, it is crucial that the sample be unbiased, so that the statistics of observables reflects the true distribution of Calabi-Yau geometries.

When constructing random sets of triangulations, there are two main considerations: we would like to sample  triangulations efficiently (i.e.~as quickly as possible) and fairly (i.e.~sampled from a uniform distribution). Unfortunately, these two goals are at odds with each other. Sampling triangulations without introducing any biases results in significant computational expense. Consequently, \texttt{CYTools} contains two separate methods to generate random triangulations: \mintinline{python}{random_triangulations_fast()} and \mintinline{python}{random_triangulations_fair()}. We will describe the associated algorithms in detail in \S\ref{sec:methods_triangulations}.

Both of these functions return a generator object, and both have a variety of parameters that can be tweaked. By default they try to find suitable parameters for the polytope being used, but it is recommended to tune the parameters by hand, especially with the fair generator, to achieve a balance between speed and the fairness of the sampling, as the default parameters tend to err on the side of fairness over speed.

Let us look at a simple example that uses these random triangulation generators.
\begin{minted}[
bgcolor=light-gray,
fontsize=\footnotesize,
escapeinside=||
]{python}
>>> p = Polytope([[1,0,0,0],[0,1,0,0],[0,0,1,0],[0,0,0,1],[-1,-1,-1,-1]]).dual()
>>> fast_triangs = p.random_triangulations_fast(N=5)
>>> for t in fast_triangs: # This loop completes very quickly
>>>     print(t)
|A fine, regular, star triangulation of a 4-dimensional point configuration with
106 points in ZZ\textsuperscript{$\wedge$}4|
|** repeats 4 more times **|
>>> fair_triangs = p.random_triangulations_fair(N=5)
>>> for t in fair_triangs: # This time it takes a long time
>>>     print(t)
|A fine, regular, star triangulation of a 4-dimensional point configuration with
106 points in ZZ\textsuperscript{$\wedge$}4|
|** repeats 4 more times **|
\end{minted}
As always, more details about these functions and their parameters can be found at the documentation website \cite{cytools-website}.

\subsection{Toric varieties}

Let us now construct a toric variety using the triangulation of a polytope. When the triangulation is star, we can interpret the maximal simplices as defining the cones of a toric fan. The corresponding toric variety can then be obtained with the \mintinline{python}{get_toric_variety} function. Here is a simple example.
\begin{minted}[
bgcolor=light-gray,
fontsize=\footnotesize,
escapeinside=||
]{python}
>>> p = Polytope([[1,0,0,0],[0,1,0,0],[0,0,1,0],[0,0,0,1],[-1,-1,-6,-9]])
>>> t = p.triangulate()
>>> v = t.get_toric_variety()
>>> print(v)
|A simplicial compact 4-dimensional toric variety with 9 affine patches|
\end{minted}
The \mintinline{python}{ToricVariety} class offers a variety of functions to compute properties of the toric variety, such as intersection numbers, Mori and K\"ahler cones, and the Stanley-Reisner ideal. We showcase many of these functions in \S\ref{sec:examples}, and refer the reader to the website for a full list.

\subsection{Calabi-Yau manifolds}

Now we can construct a Calabi-Yau hypersurface, starting from the \mintinline{python}{Triangulation} or \mintinline{python}{ToricVariety} objects.
\begin{minted}[
bgcolor=light-gray,
fontsize=\footnotesize,
escapeinside=||
]{python}
>>> p = Polytope([[1,0,0,0],[0,1,0,0],[0,0,1,0],[0,0,0,1],[-1,-1,-6,-9]])
>>> t = p.triangulate()
>>> v = t.get_toric_variety()
>>> cy = v.get_cy()
>>> cy = t.get_cy() # equivalent to the above line, but more conveninent
>>> print(cy)
|A Calabi-Yau 3-fold hypersurface with h11=2 and h21=272 in a 4-dimensional
toric variety|
\end{minted}

Again, the \mintinline{python}{Calabi-Yau} class offers a variety of functions, many of which we showcase in \S\ref{sec:examples}. However, let us mention one here since the examples below do not properly show its usefulness. It is not surprising that multiple triangulations can give rise to the same Calabi-Yau hypersurface: in particular, triangulations with identical restrictions to codimension-2 faces yield equivalent Calabi-Yaus. Moreover, automorphisms of the polytope also allow apparently different triangulations to be identified. To remove these redundancies from the analysis, \texttt{CYTools} incorporates a function called \mintinline{python}{is_trivially_equivalent} that checks for this ``trivial'' equivalence. We call the above equivalences trivial because they can be checked from triangulation and polytope data, whereas a more general equivalence check involves verifying the conditions of Wall's theorem \cite{Wall1966}, which is generally extremely difficult. Let us see how this equivalence check works in a simple example.

\begin{minted}[
bgcolor=light-gray,
fontsize=\footnotesize,
escapeinside=||
]{python}
>>> p = Polytope([[1,0,0,0],[0,1,0,0],[0,0,1,0],[0,0,0,1],
>>>               [-1,-2,-1,-1],[-2,-1,-1,-1]])
>>> all_triangs = p.all_triangulations()
>>> t1 = next(all_triangs)
>>> t2 = next(all_triangs)
>>> t1 == t2 # verify that the triangulations are different
|False|
>>> cy1 = t1.get_cy()
>>> cy2 = t2.get_cy()
>>> cy1.is_trivially_equivalent(cy2)
|True|
\end{minted}
\texttt{CYTools} automatically mods out by the above equivalence when constructing \emph{sets} of \mintinline{python}{CalabiYau} objects,
as occurs when using \texttt{\{\}} in Python.
Let us look at an example.
\begin{minted}[
bgcolor=light-gray,
fontsize=\footnotesize,
escapeinside=||
]{python}
>>> p = Polytope([[-1,0,0,0],[-1,1,0,0],[-1,0,1,0],[2,-1,0,-1],[2,0,-1,-1],
>>>               [2,-1,-1,-1],[-1,0,0,1],[-1,1,0,1],[-1,0,1,1]])
>>> all_cys = [t.get_cy() for t in p.all_triangulations()]
>>> cys_not_triv_eq = set(all_cys)
>>> # An equivalent way to directly construct this set is as follows
>>> cys_not_triv_eq = {t.get_cy() for t in p.all_triangulations()}
>>> print(len(all_cys), len(cys_not_triv_eq))
|102 5|
\end{minted}
We see that while there are a total of 102 fine, regular, star triangulations, there are only 5 Calabi-Yau hypersurfaces that are not identical via  trivial equivalence.
It remains possible that some or all of these 5 are topologically equivalent in a nontrivial way, but this is difficult to check with version 1.0 of \texttt{CYTools}.  We hope to provide capability in this direction in a future release.

There are many more functions included in the \mintinline{python}{CalabiYau} class. We will cover some of them in \S\ref{sec:examples}, but we refer the reader to the website for a comprehensive list.
\newpage

\section{Using \texttt{CYTools}: Illustrative Examples}\label{sec:examples}

In this section, we will demonstrate how to use \texttt{CYTools} to construct and analyze Calabi-Yau hypersurfaces in concrete examples. A similar demonstration can be found in the tutorial hosted on our website \cite{cytools-website}, where we also provide a Jupyter notebook in which you can follow along.

\subsection{The hypersurface in $\mathbb{P}_{[1,1,1,6,9]}$}

We will start by studying a simple Calabi-Yau with $h^{1,1}=2$, namely the anticanonical hypersurface in the weighted projective space $\mathbb{P}_{[1,1,1,6,9]}$.

\subsubsection{Polytope data}

We first import the \mintinline{python}{Polytope} class from the \texttt{CYTools} package.
\begin{minted}[
bgcolor=light-gray,
fontsize=\footnotesize,
escapeinside=||
]{python}
>>> from cytools import Polytope
\end{minted}
We define a \mintinline{python}{Polytope} object by specifying its vertices.
\begin{minted}[
bgcolor=light-gray,
fontsize=\footnotesize,
escapeinside=||
]{python}
>>> vertices = [[1,0,0,0],[0,1,0,0],[0,0,1,0],[0,0,0,1],[-1,-1,-6,-9]]
>>> p = Polytope(vertices)
>>> print(p)
|A 4-dimensional reflexive lattice polytope in ZZ\textsuperscript{$\wedge$}4|
\end{minted}
Next, we make sure that the resulting polytope is 4-dimensional,
\begin{minted}[
bgcolor=light-gray,
fontsize=\footnotesize,
escapeinside=||
]{python}
>>> p.dimension()
|4|
\end{minted}
and is reflexive,
\begin{minted}[
bgcolor=light-gray,
fontsize=\footnotesize,
escapeinside=||
]{python}
>>> p.is_reflexive()
|True|
\end{minted}
We can obtain the lattice points of the polytope.
\begin{minted}[
bgcolor=light-gray,
fontsize=\footnotesize,
escapeinside=||
]{python}
>>> p.points()
|array([[ 0,  0,  0,  0],
       [-1, -1, -6, -9],
       [ 0,  0,  0,  1],
       [ 0,  0,  1,  0],
       [ 0,  1,  0,  0],
       [ 1,  0,  0,  0],
       [ 0,  0, -2, -3],
       [ 0,  0, -1, -2],
       [ 0,  0, -1, -1],
       [ 0,  0,  0, -1]])|
\end{minted}
The output is a NumPy array containing the coordinates of all lattice points on the polytope. We use the convention that the origin is always first,\footnote{Note that indexing in Python starts at 0.} and points strictly interior to facets are always at the end. When studying Calabi-Yau hypersurfaces, one is often interested in the lattice points that are not strictly interior to facets.
Conveniently, these can be found as follows.
\begin{minted}[
bgcolor=light-gray,
fontsize=\footnotesize,
escapeinside=||
]{python}
>>> p.points_not_interior_to_facets()
|array([[ 0,  0,  0,  0],
       [-1, -1, -6, -9],
       [ 0,  0,  0,  1],
       [ 0,  0,  1,  0],
       [ 0,  1,  0,  0],
       [ 1,  0,  0,  0],
       [ 0,  0, -2, -3]])|
\end{minted}
We notice that there are three fewer points at the end of the list: the now-omitted points were strictly interior to facets. Similarly, one can use the functions,
\begin{minted}[
bgcolor=light-gray,
fontsize=\footnotesize,
escapeinside=||
]{python}
>>> p.interior_points()
>>> p.boundary_points()
>>> p.points_interior_to_facets()
>>> p.boundary_points_not_interior_to_facets()
\end{minted}
where the function names are self-explanatory. We can construct the faces of the polytope and determine the lattice points on each face. We start with the 2-faces of our polytope.
\begin{minted}[
bgcolor=light-gray,
fontsize=\footnotesize,
escapeinside=||
]{python}
>>> p.faces(d=2)
|(A 2-dimensional face of a 4-dimensional polytope in ZZ\textsuperscript{$\wedge$}4,
 A 2-dimensional face of a 4-dimensional polytope in ZZ\textsuperscript{$\wedge$}4,
 ** omitted 6 lines **
 A 2-dimensional face of a 4-dimensional polytope in ZZ\textsuperscript{$\wedge$}4,
 A 2-dimensional face of a 4-dimensional polytope in ZZ\textsuperscript{$\wedge$}4)|
\end{minted}
This polytope has ten 2-faces. For any given face of the polytope, we can find the lattice points on that face,
\begin{minted}[
bgcolor=light-gray,
fontsize=\footnotesize,
escapeinside=||
]{python}
>>> two_face = p.faces(d=2)[0]
>>> two_face.points()
|array([[-1, -1, -6, -9],
       [ 0,  0,  0,  1],
       [ 0,  0,  1,  0]])|
\end{minted}
and construct the faces of the face,
\begin{minted}[
bgcolor=light-gray,
fontsize=\footnotesize,
escapeinside=||
]{python}
>>> two_face.faces(d=1)
|(A 1-dimensional face of a 4-dimensional polytope in ZZ\textsuperscript{$\wedge$}4,
 A 1-dimensional face of a 4-dimensional polytope in ZZ\textsuperscript{$\wedge$}4,
 A 1-dimensional face of a 4-dimensional polytope in ZZ\textsuperscript{$\wedge$}4)|
\end{minted}
Some relevant data of the Calabi-Yau hypersurface can be computed directly from the polytope. For example, the Hodge numbers and Euler characteristic can be computed as follows.
\begin{minted}[
bgcolor=light-gray,
fontsize=\footnotesize,
escapeinside=||
]{python}
>>> p.h11(lattice="N")
|2|
>>> p.h21(lattice="N")
|272|
>>> p.chi(lattice="N")
|-540|
\end{minted}
Note that one must specify the lattice on which the polytope lives. Setting \mintinline{python}{lattice="N"} or \mintinline{python}{lattice="M"} specifies that the ambient toric variety is constructed from a desingularization of the face fan or normal fan of the polytope, respectively. We can check whether the polytope is favorable.
\begin{minted}[
bgcolor=light-gray,
fontsize=\footnotesize,
escapeinside=||
]{python}
>>> p.is_favorable(lattice="N")
True
\end{minted}
We can verify that switching lattices we obtain the data of the mirror Calabi-Yau.
\begin{minted}[
bgcolor=light-gray,
fontsize=\footnotesize,
escapeinside=||
]{python}
>>> p.h11(lattice="M")
|272|
>>> p.h21(lattice="M")
|2|
>>> p.chi(lattice="M")
|540|
\end{minted}
We can construct the dual polytope,
\begin{minted}[
bgcolor=light-gray,
fontsize=\footnotesize,
escapeinside=||
]{python}
>>> p.dual_polytope()
|A 4-dimensional reflexive lattice polytope in ZZ\textsuperscript{$\wedge$}4|
\end{minted}
and verify that it is a 4-dimensional reflexive polytope where the Hodge numbers are switched.
\begin{minted}[
bgcolor=light-gray,
fontsize=\footnotesize,
escapeinside=||
]{python}
>>> p.dual_polytope().dimension()
|4|
>>> p.dual_polytope().is_reflexive()
|True|
>>> p.dual_polytope().h11(lattice="N")
|272|
>>> p.dual_polytope().h21(lattice="N")
|2|
\end{minted}
We can find the faces of the dual polytope just as we did with the original polytope. For example, we can find the 1-faces.
\begin{minted}[
bgcolor=light-gray,
fontsize=\footnotesize,
escapeinside=||
]{python}
>>> p.dual_polytope().faces(d=1)
|(A 1-dimensional face of a 4-dimensional polytope in ZZ\textsuperscript{$\wedge$}4,
 A 1-dimensional face of a 4-dimensional polytope in ZZ\textsuperscript{$\wedge$}4,
 ** omitted 6 lines **
 A 1-dimensional face of a 4-dimensional polytope in ZZ\textsuperscript{$\wedge$}4,
 A 1-dimensional face of a 4-dimensional polytope in ZZ\textsuperscript{$\wedge$}4)|
\end{minted}
There are ten 1-faces of the dual polytope, as expected. The lists of faces are ordered such that the $n^{\mathrm{th}}$ 2-face of the original polytope is dual to the $n^{\mathrm{th}}$ 1-face of the dual polytope.
\begin{minted}[
bgcolor=light-gray,
fontsize=\footnotesize,
escapeinside=||
]{python}
>>> n = 0
>>> p.dual_polytope().faces(d=1)[n].dual_face().points()
|array([[-1, -1, -6, -9],
       [ 0,  0,  0,  1],
       [ 0,  0,  1,  0]])|
>>> p.faces(d=2)[n].points()
|array([[-1, -1, -6, -9],
       [ 0,  0,  0,  1],
       [ 0,  0,  1,  0]])|
\end{minted}
Other important data that can be computed directly from the polytope is the GLSM charge matrix,
\begin{minted}[
bgcolor=light-gray,
fontsize=\footnotesize,
escapeinside=||
]{python}
>>> p.glsm_charge_matrix()
|array([[-18,   1,   9,   6,   1,   1,   0],
       [ -6,   0,   3,   2,   0,   0,   1]])|
\end{minted}
where the first column corresponds to the canonical divisor, and the remaining columns to the prime toric divisors $D^i$. By default, divisors corresponding to points strictly interior to facets are not included, as they do not intersect a generic Calabi-Yau hypersurface.

It is often useful to pick a basis of $h^{1,1}$ divisors that span $H_4(X, \mathbb{Z})$. We do this by picking a subset of of the prime toric divisors, such that the extra divisors can be written as integer linear combinations of the basis divisors. We can obtain such a basis as follows.
\begin{minted}[
bgcolor=light-gray,
fontsize=\footnotesize,
escapeinside=||
]{python}
>>> p.glsm_basis()
|array([5, 6])|
\end{minted}
As we can see, this function returns a list of indices indicating that divisors $D^5$ and $D^6$ form a complete basis.\footnote{We reiterate that indices start from zero. In this example, $D^1$ has GLSM charges $(1,0)$ and $D^6$ has charges $(0,1)$.} We can obtain the linear relations between the canonical and prime toric divisors as follows,
\begin{minted}[
bgcolor=light-gray,
fontsize=\footnotesize,
escapeinside=||
]{python}
>>> p.glsm_linear_relations()
|array([[ 1,  0,  0,  0,  0, 18,  6],
       [ 0,  1,  0,  0,  0, -1,  0],
       [ 0,  0,  1,  0,  0, -9, -3],
       [ 0,  0,  0,  1,  0, -6, -2],
       [ 0,  0,  0,  0,  1, -1,  0]])|
\end{minted}
where each row encodes data of a particular linear relation, while each column corresponds to a divisor. From this matrix we can see that the five divisors that are not in the basis can be written as linear combinations of $D^5$ and $D^6$. As we will see, the \mintinline{python}{ToricVariety} and \mintinline{python}{CalabiYau} classes allow picking a custom basis of divisors or curves, which are used for basis-dependent computations.

\subsubsection{Triangulation data}

Now that we have computed most relevant data that exclusively comes from the polytope, we will triangulate the polytope to obtain a toric fan.
\begin{minted}[
bgcolor=light-gray,
fontsize=\footnotesize,
escapeinside=||
]{python}
>>> t = p.triangulate()
>>> print(t)
|A fine, regular, star triangulation of a 4-dimensional point configuration with
7 points in ZZ\textsuperscript{$\wedge$}4|
\end{minted}
As we can see, by default this function returns a fine, regular, star triangulation. We can also see that by default it only used the points not interior to facets. We can verify all of this as follows.
\begin{minted}[
bgcolor=light-gray,
fontsize=\footnotesize,
escapeinside=||
]{python}
>>> t.is_fine()
|True|
>>> t.is_regular()
|True|
>>> t.is_star()
|True|
>>> t.points()
|array([[ 0,  0,  0,  0],
       [-1, -1, -6, -9],
       [ 0,  0,  0,  1],
       [ 0,  0,  1,  0],
       [ 0,  1,  0,  0],
       [ 1,  0,  0,  0],
       [ 0,  0, -2, -3]])|
\end{minted}
We can print the simplices of the triangulation,
\begin{minted}[
bgcolor=light-gray,
fontsize=\footnotesize,
escapeinside=||
]{python}
>>> t.simplices()
|array([[0, 1, 2, 3, 4],
       [0, 1, 2, 3, 5],
       [0, 1, 2, 4, 6],
       [0, 1, 2, 5, 6],
       [0, 1, 3, 4, 6],
       [0, 1, 3, 5, 6],
       [0, 2, 3, 4, 5],
       [0, 2, 4, 5, 6],
       [0, 3, 4, 5, 6]])|
\end{minted}
where each simplex is represented by the indices of its vertices.
We can also calculate the GKZ vector \cite{Gelfand1994} of the triangulation.
\begin{minted}[
bgcolor=light-gray,
fontsize=\footnotesize,
escapeinside=||
]{python}
>>> t.gkz_phi()
|array([18, 12,  9, 12, 12, 12, 15])|
\end{minted}
As a last example, we can find the triangulations that can be obtained by performing a single bistellar flip.
\begin{minted}[
bgcolor=light-gray,
fontsize=\footnotesize,
escapeinside=||
]{python}
>>> t.neighbor_triangulations()
|[A non-fine, star triangulation of a 4-dimensional point configuration with
 7 points in ZZ\textsuperscript{$\wedge$}4,
 A non-fine, non-star triangulation of a 4-dimensional point configuration with
 7 points in ZZ\textsuperscript{$\wedge$}4]|
\end{minted}

\subsubsection{Toric variety data}

Now that we have constructed a fine, regular, star triangulation, let us use it to construct the corresponding toric variety. We can do this as follows.
\begin{minted}[
bgcolor=light-gray,
fontsize=\footnotesize,
escapeinside=||
]{python}
>>> v = t.get_toric_variety()
>>> print(v)
|A simplicial compact 4-dimensional toric variety with 9 affine patches|
\end{minted}
We can see that the nine simplices of the triangulation above now correspond to affine patches of the toric variety. This is because they now correspond to the maximal cones of the toric fan that defines the variety. Let us take a look at these cones of the fan.
\begin{minted}[
bgcolor=light-gray,
fontsize=\footnotesize,
escapeinside=||
]{python}
>>> v.fan_cones()
|(A 4-dimensional rational polyhedral cone in RR\textsuperscript{$\wedge$}4 generated by 4 rays,
 A 4-dimensional rational polyhedral cone in RR\textsuperscript{$\wedge$}4 generated by 4 rays,
 A 4-dimensional rational polyhedral cone in RR\textsuperscript{$\wedge$}4 generated by 4 rays,
 A 4-dimensional rational polyhedral cone in RR\textsuperscript{$\wedge$}4 generated by 4 rays,
 A 4-dimensional rational polyhedral cone in RR\textsuperscript{$\wedge$}4 generated by 4 rays,
 A 4-dimensional rational polyhedral cone in RR\textsuperscript{$\wedge$}4 generated by 4 rays,
 A 4-dimensional rational polyhedral cone in RR\textsuperscript{$\wedge$}4 generated by 4 rays,
 A 4-dimensional rational polyhedral cone in RR\textsuperscript{$\wedge$}4 generated by 4 rays,
 A 4-dimensional rational polyhedral cone in RR\textsuperscript{$\wedge$}4 generated by 4 rays)|
\end{minted}
We can similarly find lower-dimensional cones. For example, we will look at the 1-dimensional cones, or rays, of the fan, which are in correspondence with the prime toric divisors.
\begin{minted}[
bgcolor=light-gray,
fontsize=\footnotesize,
escapeinside=||
]{python}
>>> v.fan_cones(d=1)
|(A 1-dimensional rational polyhedral cone in RR\textsuperscript{$\wedge$}4 generated by 1 rays,
 A 1-dimensional rational polyhedral cone in RR\textsuperscript{$\wedge$}4 generated by 1 rays,
 A 1-dimensional rational polyhedral cone in RR\textsuperscript{$\wedge$}4 generated by 1 rays,
 A 1-dimensional rational polyhedral cone in RR\textsuperscript{$\wedge$}4 generated by 1 rays,
 A 1-dimensional rational polyhedral cone in RR\textsuperscript{$\wedge$}4 generated by 1 rays,
 A 1-dimensional rational polyhedral cone in RR\textsuperscript{$\wedge$}4 generated by 1 rays)|
\end{minted}
Another important piece of information we obtain from the fan is the Stanley-Reisner ideal, which tells us the combinations of toric coordinates that cannot vanish at the same time.
\begin{minted}[
bgcolor=light-gray,
fontsize=\footnotesize,
escapeinside=||
]{python}
>>> v.sr_ideal()
|((1, 4, 5), (2, 3, 6))|
\end{minted}
Each tuple represents a monomial in terms of indices of toric coordinates. In this case, the Stanley-Reisner ideal is $(z_1 z_4 z_5, z_2 z_3 z_6)$, where the $z_i$ are toric coordinates.

Let us now compute the Mori cone, K\"ahler cone, and effective cone of the toric variety. These are important cones, as they give approximations of the corresponding cones for the Calabi-Yau hypersurface.
\begin{minted}[
bgcolor=light-gray,
fontsize=\footnotesize,
escapeinside=||
]{python}
>>> v.mori_cone()
|A 2-dimensional rational polyhedral cone in RR\textsuperscript{$\wedge$}7 generated by 3 rays|
>>> v.mori_cone(in_basis=True)
|A 2-dimensional rational polyhedral cone in RR\textsuperscript{$\wedge$}2 generated by 3 rays|
>>> v.kahler_cone()
|A rational polyhedral cone in RR\textsuperscript{$\wedge$}2 defined by 3 hyperplanes|
>>> v.effective_cone()
|A 2-dimensional rational polyhedral cone in RR\textsuperscript{$\wedge$}2 generated by 6 rays|
\end{minted}
Note that cones in \texttt{CYTools} can be defined in two ways: by specifying a set of rays, or by specifying the defining hyperplanes.

The Mori cone can be interpreted as a cone living in an $(h^{1,1}+5)$-dimensional lattice, or in an $h^{1,1}$-dimensional lattice after a choice of basis, while the K\"ahler and effective cones always require a choice of basis. We will discuss basis choices below, but for now let us conclude by computing intersection numbers of the variety with and without a choice of basis.
\begin{minted}[
bgcolor=light-gray,
fontsize=\footnotesize,
escapeinside=||
]{python}
>>> v.intersection_numbers()
|\{(1, 2, 3, 4): 1.0,
 (1, 2, 3, 5): 1.0,
** omitted 117 lines **
 (0, 0, 0, 5): -108,
 (0, 0, 0, 0): 1944\}|
>>> v.intersection_numbers(in_basis=True)
|\{(0, 0, 1, 1): 0.16666666666666757,
 (0, 1, 1, 1): -1.0000000000000027,
 (1, 1, 1, 1): 4.500000000000011\}|
\end{minted}
The output of this function is a dictionary that contains key-value pairs for the non-zero intersection numbers. The keys consist of sorted indices of the divisors, and the values are the corresponding intersection numbers. When we compute intersection numbers without a basis choice, indices range from zero to $h^{1,1}+5$, corresponding to the canonical and prime toric divisors. From the output we can see that, for example, $\kappa_{1234}=1$. When using a choice of basis, the indices range from zero to $h^{1,1}-1$, corresponding to the divisors in the basis. Since we previously saw that the default basis was $\{\hat{D}^5,\hat{D}^6\}$, this output tells us that $\#\hat{D}^5\cap \hat{D}^6\cap \hat{D}^6\cap \hat{D}^6=-1$.
It is important to note that intersection numbers are computed as (double-precision) floating-point numbers. This provides a performance advantage since the fastest linear solvers work with floats. For most computations, double-precision floats offer enough precision to accurately represent the rational numbers that arise.

The intersection numbers of smooth Calabi-Yau hypersurfaces must be integers, and a verification is always performed to make sure that the intersection numbers computed in \texttt{CYTools} do not deviate from integrality by a significant amount. We discuss the computation of intersection numbers in \S\ref{sec:methods_intnums}.

\subsubsection{Calabi-Yau data}

Finally, let us get to the object of most interest. We can construct a \mintinline{python}{CalabiYau} object from the \mintinline{python}{ToricVariety} we constructed above, but we can also get it directly from the \mintinline{python}{Triangulation} object.
\begin{minted}[
bgcolor=light-gray,
fontsize=\footnotesize,
escapeinside=||
]{python}
>>> cy = t.get_cy()
>>> print(cy)
|A Calabi-Yau 3-fold hypersurface with h11=2 and h21=272 in a 4-dimensional
toric variety|
\end{minted}
First, let us make sure that the generic hypersurface is smooth.
\begin{minted}[
bgcolor=light-gray,
fontsize=\footnotesize,
escapeinside=||
]{python}
>>> cy.is_smooth()
|True|
\end{minted}
We can calculate the intersection numbers of the hypersurface just as we did with the toric variety.
\begin{minted}[
bgcolor=light-gray,
fontsize=\footnotesize,
escapeinside=||
]{python}
>>> cy.intersection_numbers()
|\{(1, 2, 3): 18,
 (2, 3, 4): 18,
** omitted 52 lines **
 (0, 0, 5): 108,
 (0, 0, 0): -1944\}|
>>> cy.intersection_numbers(in_basis=True)
|\{(0, 1, 1): -3, (1, 1, 1): 9, (0, 0, 1): 1\}|
\end{minted}
Similar to what we saw for toric varieties, the indices correspond to canonical and prime toric divisors, or divisors in the basis, depending on whether we set \mintinline{python}{in_basis=True}.

This is a good opportunity to talk about how divisor bases work in \texttt{CYTools}.  By default, the \mintinline{python}{ToricVariety} and \mintinline{python}{CalabiYau} classes use the basis that the \mintinline{python}{glsm_basis} function of the \mintinline{python}{Polytope} class returns. We can verify this as follows.
\begin{minted}[
bgcolor=light-gray,
fontsize=\footnotesize,
escapeinside=||
]{python}
>>> cy.divisor_basis()
|array([5, 6])|
\end{minted}
As expected, we see that $D^5$ and $D^6$ are used as the basis divisors.\footnote{Given a set of basis divisors $D_i$, \texttt{CYTools} will also use a dual basis of curves $C_j$, so that $D_i\cdot C_j=\delta_{ij}$.}

We also have the option of choosing a custom basis, but with the requirement of it being integral.  In the above example, $D^1$ and $D^6$ can be picked as our basis. To set this we do the following.
\begin{minted}[
bgcolor=light-gray,
fontsize=\footnotesize,
escapeinside=||
]{python}
>>> cy.set_divisor_basis([1, 6])
>>> cy.divisor_basis()
|array([1, 6])|
>>> cy.curve_basis()
|array([1, 6])|
\end{minted}
Now all the computations that are basis-dependent will be done with this choice of basis. Similarly, a new basis of curves can be set, and the divisor basis will also be updated. This case is particularly simple because the new basis is equivalent to the old one, but in more elaborate examples the specification of a basis can be useful. Furthermore, it is also possible to set a basis of divisors that corresponds to a general linear combination\footnote{As noted above, only integral bases are allowed.} of prime toric divisors.
We refer the reader to the website for more details on these more advanced use cases.

Now that we have seen how bases work, let us move on and compute the second Chern class. We represent this data as a vector of the integrals of the second Chern class either over the canonical and prime toric divisors, or over the current basis of divisors.
\begin{minted}[
bgcolor=light-gray,
fontsize=\footnotesize,
escapeinside=||
]{python}
>>> cy.second_chern_class()
|array([-612,   36,  306,  204,   36,   36,   -6])|
>>> cy.second_chern_class(in_basis=True)
|array([36, -6])|
\end{minted}

The cone of effective divisors of the ambient variety is generated by the $h^{1,1}+4$ prime toric divisors $D^i$. We can express each of these divisors in terms of the $h^{1,1}$ basis divisors. The resulting vectors define a solid $h^{1,1}$-dimensional cone, which gives an inner approximation of the effective cone of the Calabi-Yau hypersurface. We call this approximation the toric effective cone.
\begin{minted}[
bgcolor=light-gray,
fontsize=\footnotesize,
escapeinside=||
]{python}
>>> eff_cone = cy.toric_effective_cone()
>>> print(eff_cone)
|A 2-dimensional rational polyhedral cone in RR\textsuperscript{$\wedge$}2 generated by 6 rays|
\end{minted}
The presentation of the effective cone is basis dependent, so changing to a new basis will result in a different set of rays.

Next we construct the toric approximation of the Mori cone,
\begin{minted}[
bgcolor=light-gray,
fontsize=\footnotesize,
escapeinside=||
]{python}
>>> mori_cone = cy.toric_mori_cone()
>>> print(mori_cone)
|A 2-dimensional rational polyhedral cone in RR\textsuperscript{$\wedge$}7 generated by 3 rays|
\end{minted}
and compute its generating rays.
\begin{minted}[
bgcolor=light-gray,
fontsize=\footnotesize,
escapeinside=||
]{python}
>>> mori_cone.rays()
|array([[  0,   1,   0,   0,   1,   1,  -3],
       [-18,   1,   9,   6,   1,   1,   0],
       [ -6,   0,   3,   2,   0,   0,   1]])|
\end{minted}
Each row corresponds to an effective curve $C_\alpha$, and the entries along columns correspond to the intersection numbers with the canonical and prime toric divisors. We can compute the extremal rays as follows.
\begin{minted}[
bgcolor=light-gray,
fontsize=\footnotesize,
escapeinside=||
]{python}
>>> mori_cone.extremal_rays()
|array([[ 0,  1,  0,  0,  1,  1, -3],
       [-6,  0,  3,  2,  0,  0,  1]])|
\end{minted}
It is often useful to view the Mori cone as living in $H_2(X, \mathbb{Z})$. We do this by expressing it in the chosen basis of curves, which is dual to the basis of divisors.
\begin{minted}[
bgcolor=light-gray,
fontsize=\footnotesize,
escapeinside=||
]{python}
>>> mori_cone = cy.toric_mori_cone(in_basis=True)
>>> mori_cone.rays()
|array([[ 1, -3],
       [ 0,  1]])|
\end{minted}
We can construct the toric part of the K\"ahler cone,
\begin{minted}[
bgcolor=light-gray,
fontsize=\footnotesize,
escapeinside=||
]{python}
>>> kahler_cone = cy.toric_kahler_cone()
>>> print(kahler)
|A rational polyhedral cone in RR\textsuperscript{$\wedge$}2 defined by 2 hyperplanes|
\end{minted}
which is equivalent to taking the dual cone of the toric Mori cone in a specified basis.
We can find the tip of the stretched K\"ahler cone, i.e.~the shortest\footnote{Length is measured with the Euclidean metric in the working basis.  See \cite{Demirtas:2018akl} for the motivation for considering this point in the moduli space.} vector in the K\"ahler cone for which the generators of the Mori cone have volume at least $c$.
\begin{minted}[
bgcolor=light-gray,
fontsize=\footnotesize,
escapeinside=||
]{python}
>>> tip = kahler_cone.tip_of_stretched_cone(c=1)
>>> tip
|array([4., 1.])|
\end{minted}
Next, let us compute the volume $\mathcal{V}$ of the Calabi-Yau hypersurface, the volumes of the prime toric divisors $\tau^{i}$, and the volumes of the generators of the Mori cone, all at
this point in the K\"ahler cone.
\begin{minted}[
bgcolor=light-gray,
fontsize=\footnotesize,
escapeinside=||
]{python}
>>> cy.compute_cy_volume(tip)
|3.4999999999999973|
>>> cy.compute_divisor_volumes(tip)
|array([ 2.5, 24. , 16. ,  2.5,  2.5,  0.5])|
>>> cy.compute_curve_volumes(tip)
array([1., 1.])
\end{minted}
Finally, we define the K\"ahler metric $K_{ij}=\frac{\partial \mathcal{K}}{\partial \tau^i \partial \tau^j}$, where $\mathcal{K} = - \log (2 \mathcal{V})$. Its inverse  $(K_{ij})^{-1}$ can be expressed in closed form and computed efficiently.
\begin{minted}[
bgcolor=light-gray,
fontsize=\footnotesize,
escapeinside=||
]{python}
>>> cy.compute_inverse_kahler_metric(tip)
|array([[11., -9.],
       [-9., 43.]])|
\end{minted}
This concludes the walkthrough of one of the simplest examples. There is functionality that we did not cover in the interest of space, but all the functionality is documented on the website \cite{cytools-website}.

\subsection{A Calabi-Yau with $h^{1,1}=491$}

Let us now consider a Calabi-Yau hypersurface arising from the largest polytope in the Kreuzer-Skarke database. The computations work in the same way as in the previous examples, except with much larger output data. Nevertheless, this example highlights the capabilities of \texttt{CYTools}, as it proves that even the most intricate Calabi-Yau manifolds can be easily analyzed.

Let us start by constructing the polytope. We can construct it directly from the vertices, but we can instead use the \mintinline{python}{fetch_polytopes} function if we do not have them memorized.
\begin{minted}[
bgcolor=light-gray,
fontsize=\footnotesize,
escapeinside=||
]{python}
>>> from cytools import fetch_polytopes
>>> p = next(fetch_polytopes(h11=491, lattice="N"))
>>> p.vertices()
|array([[  1,   0,   0,   0],
       [  0,   1,   0,   0],
       [  0,   0,   1,   0],
       [ 21,  28,  36,  42],
       [-63, -56, -48, -42]])|
\end{minted}

We now construct a triangulation and Calabi-Yau hypersurface as before.
\begin{minted}[
bgcolor=light-gray,
fontsize=\footnotesize,
escapeinside=||
]{python}
>>> t = p.triangulate()
>>> cy = t.get_cy()
>>> print(cy)
|A Calabi-Yau 3-fold hypersurface with h11=491 and h21=11 in a 4-dimensional
toric variety|
\end{minted}

The GLSM charge matrix will now be a very large matrix, but it can still be computed efficiently.
\begin{minted}[
bgcolor=light-gray,
fontsize=\footnotesize,
escapeinside=||
]{python}
>>> cy.glsm_charge_matrix()
|array([[-330,    1,    0, ...,    0,    0,    0],
       [   6,    0,    1, ...,    0,    0,    0],
       [ 174,    0,    0, ...,    0,    0,    0],
       ...,
       [ 132,    0,    0, ...,    1,    0,    0],
       [ 138,    0,    0, ...,    0,    1,    0],
       [ 144,    0,    0, ...,    0,    0,    1]])|
\end{minted}

Similarly, the Mori cone has a large number of generators and lives in a high-dimensional lattice, but it can easily be computed.
\begin{minted}[
bgcolor=light-gray,
fontsize=\footnotesize,
escapeinside=||
]{python}
>>> cy.toric_mori_cone()
|A 491-dimensional rational polyhedral cone in RR\textsuperscript{$\wedge$}496 generated by 3696 rays|
\end{minted}

Crucially, even hard computations like intersection numbers only take about a second.
\begin{minted}[
bgcolor=light-gray,
fontsize=\footnotesize,
escapeinside=||
]{python}
>>> cy.intersection_numbers()
|\{(1, 14, 28): 1,
 (1, 6, 28): 1,
** omitted many lines**|
>>> len(cy.intersection_numbers()) # the number of non-zero intersection numbers
|4643|
\end{minted}

Lastly, let us find the tip of the stretched K\"ahler cone, and find the volume of the Calabi-Yau at that point.
\begin{minted}[
bgcolor=light-gray,
fontsize=\footnotesize,
escapeinside=||
]{python}
>>> tip = cy.toric_kahler_cone().tip_of_stretched_cone(c=1)
>>> cy.compute_cy_volume(tip)
|1729010461108.0|
\end{minted}

\subsection{An example scan} \label{sec:exscan}

Finally, we show a short Python script that efficiently constructs an ensemble of geometries by scanning over the Kreuzer-Skarke database.
For this illustration, we perform a scan similar to that presented in \cite{Demirtas:2018akl}: in particular, we produce plots analogous to figures 1, 3, 4, and 8(a) of \cite{Demirtas:2018akl}. We use a much smaller set of polytopes so that the code only takes a few minutes to run, but improve on the original scan by sampling multiple triangulations per polytope.
Here we only demonstrate a fast (biased) sampling to keep the run time to a few minutes, but this code is easily adapted to perform fairer samplings of triangulations. This code can be found in the tutorial section on the website \cite{cytools-website}.

\begin{minted}[
bgcolor=light-gray,
fontsize=\footnotesize,
escapeinside=||
]{python}
# We start by importing fetch_polytopes,
# a plotting package, and numpy
from cytools import fetch_polytopes
import matplotlib.pyplot as plt
import numpy as np

# These are the settings for the scan.
# We scan h11=2,3,4,5,10,15,...,100
# For each h11 we take 10 polytopes,
# and 5 random triangulations for each polytope
h11s = [2,3,4] + list(range(5,105,5))
n_polys = 10
n_triangs = 5

# These are the lists where we will save the data
h11_list = []
nonzerointnums = []
costhetamin = []
dmins = []
Xvols = []

for h11 in h11s:
    print(f"Processing h11={h11}", end="\r")
    for p in fetch_polytopes(h11=h11, lattice="N",
                             favorable=True, limit=n_polys):
        # Here we take a random set of triangulations by picking random heights.
        # We use the random_triangulations_fast function with max_retries=5 so
        # that the generation doesn't take too long. However, this will not
        # generate a fair sampling of the triangulations of the polytope.
        # For a fair sampling one should use the random_triangulations_fair
        # function, which is much slower.
        for t in p.random_triangulations_fast(N=n_triangs, max_retries=5):
            cy = t.get_cy()
            h11_list.append(h11)
            nonzerointnums.append(len(cy.intersection_numbers(in_basis=True)))
            mori_rays = cy.toric_mori_cone(in_basis=True).rays()
            mori_rays_norms = np.linalg.norm(mori_rays, axis=1)
            n_mori_rays = len(mori_rays)
            costhetamin.append(min(
                mori_rays[i].dot(mori_rays[j])
                    /(mori_rays_norms[i]*mori_rays_norms[j])
                for i in range(n_mori_rays) for j in range(i+1,n_mori_rays)))
            tip = cy.toric_kahler_cone().tip_of_stretched_cone(1)
            dmins.append(np.log10(np.linalg.norm(tip)))
            Xvols.append(np.log10(cy.compute_cy_volume(tip)))
print("Finished processing all h11s!")
print(f"Scanned through {len(h11_list)} CY hypersurfaces.")

# We plot the data using matplotlib.
# If you are not familiar with this package, you can find tutorials and
# documentation at https://matplotlib.org/

xdata = [h11_list]*3 + [np.log10(h11_list)]
ydata = [nonzerointnums, costhetamin, dmins, Xvols]
xlabels = [r"$h^{1,1}$"]*3 + [r"log${}_{10}(h^{1,1})$"]
ylabels = [r"# nonzero $\kappa_{ijk}$", r"$\cos(\theta_{min})$",
           r"log${}_{10}(d_{min})$", r"log${}_{10}(\mathcal{V})$"]
fig, ax0 = plt.subplots(2, 2, figsize=(15,12))

for i,d in enumerate(ydata):
    ax = plt.subplot(221+i)
    ax.scatter(xdata[i], ydata[i], s=10)
    plt.xlabel(xlabels[i], size=20)
    plt.ylabel(ylabels[i], size=20)
    plt.tick_params(labelsize=15, width=2, length=5)

plt.subplots_adjust(wspace=0.3, hspace=0.22)
\end{minted}

\begin{figure}[!ht]
    \centering
    \includegraphics[width=\textwidth]{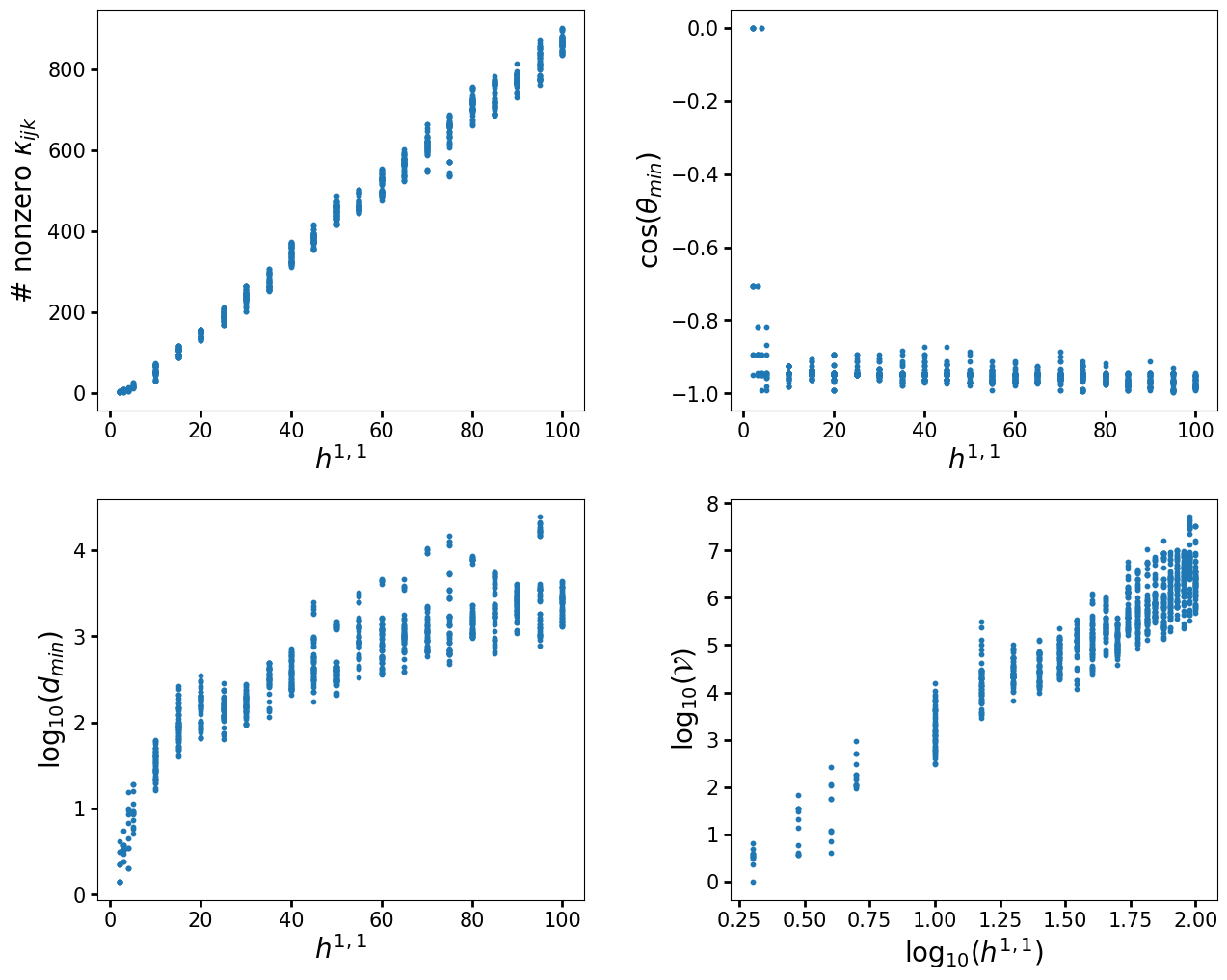}
	\caption{The resulting plots from the example code.  Definitions and explanations can be found in \cite{Demirtas:2018akl}.}
	\label{fig:scan_plots}
\end{figure}

After running the code, we obtain the plots shown in Fig.~\ref{fig:scan_plots}.
Comparing with the plots from the original paper  \cite{Demirtas:2018akl}, we can see that the general trends match, but there are some slight variations. This is mainly because the triangulations used here are constructed using heights similar to those of the Delaunay triangulations, while in the original paper the triangulations were constructed using \texttt{TOPCOM}, which uses a push-pulling algorithm. These two approaches yield triangulations with qualitatively different properties, and thus result in Calabi-Yau hypersurfaces with slightly different properties. On the website we show how to obtain the triangulations with \texttt{TOPCOM} to reproduce the original results.\footnote{The polytopes retrieved by \mintinline{python}{fetch_polytopes} are not randomly sampled, and are instead retrieved in the order in which they are listed in the Kreuzer-Skarke database. When $h^{1,1}$ is specified but $h^{2,1}$ is not, the polytopes are retrieved in order of ascending $h^{2,1}$.}

The above example shows the power of \texttt{CYTools}. The original paper required very significant effort to assemble the code for the analysis, as it required downloading the Kreuzer-Skarke database, performing some computations in \texttt{SageMath}, then performing some extra computations in \texttt{Mathematica}, and finally using a variety of scripts to gather together all the data. Now, anyone can perform the full analysis with a few lines of code in \texttt{CYTools}. Although we only took a small number of polytopes in this example, one could easily increase the range of the scan and even surpass the statistics of the original paper by running the computation on a standard laptop overnight.

\section{Methods and Algorithms}\label{sec:methods}
In this section, we describe the algorithms and external software packages utilized by \texttt{CYTools}.

\subsection{Polytopes}

Polytopes are handled in \texttt{CYTools} by the \mintinline{python}{Polytope} class, which provides a variety of functionality to study lattice polytopes. There are three available backends
that can find the hyperplane representation from a given vertex representation: \texttt{PPL} \cite{BagnaraHZ08SCP}, \texttt{QHull} \cite{Barber96}, and \texttt{PALP} \cite{Kreuzer:2002uu}.  The default backend is \texttt{PPL}, as it is slightly faster than the other two in typical scenarios, but the others can be chosen by setting the \mintinline{python}{backend} argument to \mintinline{python}{"qhull"} or \mintinline{python}{"palp"}.
While \texttt{PALP} provides further functionality to handle lattice polytopes, the other two do not.  Nevertheless, we still do not use \texttt{PALP} in most cases, as we have implemented faster algorithms to perform some computations on 4-dimensional polytopes. We now describe the two main algorithms.

\subsubsection{Lattice points}\label{sec:lattice-points}

One of the most basic computations for a lattice polytope is finding all of its lattice points. Even though this sounds like an easy task, it is fairly challenging to do it efficiently. \texttt{CYTools} comes with two different ways of performing this computation: a native computation in Python, and one using \texttt{PALP} \cite{Kreuzer:2002uu} as the backend. We will start by describing the native algorithm, which is based on Volker Braun's code written for the \mintinline{python}{LatticePolytope_PPL} class \cite{latticepolytope-ppl} of the \texttt{SageMath} software package \cite{sagemath}. The procedure is described in Algorithm \ref{algo:lattice_points}. It is assumed without loss of generality that the polytope is full-dimensional because otherwise we can simply consider it as living on the affine lattice that it spans.

\vspace{3mm}
\setlength{\algomargin}{0em}
\begin{algorithm}[H]
    \begin{enumerate}[labelindent=0pt]
        \item\label{algo:step_lll} Perform the Lenstra–Lenstra–Lov\'asz (LLL) lattice basis reduction algorithm \cite{LLL} on the vertices $v_i$ of the polytope to obtain new vertices $v_i'$.
        \item Find the smallest rectangular box containing the vertices $v_i'$. Let the intervals defining the box be $[b_{i,\text{min}},b_{i,\text{max}}]$ with $i=1,\cdots,n$ where $n$ is the dimension of the lattice. The coordinates are reordered so that the dimensions of the box are in decreasing size $(b_{1,\text{max}}-b_{1,\text{min}})\geq\cdots\geq(b_{n,\text{max}}-b_{n,\text{min}})$.
        \item Fix the coordinates $x_i=b_{i,\text{min}}$ with $i=2,\cdots,n$.
        \item\label{algo:step_find_lattice_points} Find the maximum and minimum values $x_{1,\text{max}},x_{1,\text{min}}$ of $x_1$ such that points with the previously fixed coordinates are contained in the polytope. The rest of the points with values of $x_1$ in between these will also be contained
        \item Increase the coordinates $x_i$, $i=2,\cdots,n$ lexicographically, performing step \ref{algo:step_find_lattice_points} after each increment, thus finding all lattice points of the polytope.
        \item Perform the inverse LLL transformation on the points to return to the original presentation of the polytope.
    \end{enumerate}
    \caption{Find the lattice points of $\Delta$.}
    \label{algo:lattice_points}
\end{algorithm}
\vspace{3mm}

The addition of step \ref{algo:step_lll} provides a significant advantage over Braun's original algorithm, since it generally results in an appreciable reduction of the size of the rectangular bounding box, thus speeding up the later steps. This allows this Python algorithm to have a similar level of performance to \texttt{PALP}, which is written in the C programming language, at least when $\dim(\Delta)\leq 4$. In fact, when  Algorithm \ref{algo:lattice_points} is implemented in a faster programming language such as Julia, it can considerably outperform \texttt{PALP}.

As previously mentioned, the alternative method to compute the lattice points of the polytope is by using \texttt{PALP}. When the \mintinline{python}{Polytope} class uses \texttt{PPL} or \texttt{Qhull} as the backend, then Algorithm \ref{algo:lattice_points} is used, and when it uses \texttt{PALP} as the backend then \texttt{PALP} is also used for this computation. Earlier we noted that the \mintinline{python}{Polytope} class uses \texttt{PPL} as the default backend for dimensions less than or equal to 4. This is because, as dimension increases, the ratio of the volume of a generic polytope to the volume of its bounding box decreases, so Algorithm \ref{algo:lattice_points} becomes less efficient and \texttt{PALP} surpasses it.

Lastly, we remark that when \texttt{CYTools} finds the lattice points of a polytope, it also checks which of the hyperplane inequalities defining the polytope each point saturates, storing the point along with a set of indices of the saturated inequalities. This will be a key step for the computation of the faces of the polytope, as we will now describe.

\subsubsection{Faces}\label{sec:faces}

We now explain how we compute the set of faces of a polytope. The algorithm was developed from scratch with speed as the main goal, but the key idea is identical to well-known backtracking algorithms for face computations (see e.g.~\cite{FUKUDA19971}). The data required for the computation are the hyperplane inequalities defining the polytope, and the set of vertices along with the inequalities that they saturate. In the algorithm we assume that the polytope $\Delta$ is $n$-dimensional, is defined by $m$ hyperplane inequalities (i.e. it has $m$ facets), and is full-dimensional (i.e. there are no hyperplane equalities). If it is not full-dimensional we can consider it as living in the affine lattice that it spans.  We index the hyperplane inequalities by $1,\cdots,m$ and denote the set of inequalities saturated by a point $p_i$ as $\mathcal{S}(p_i)$. Lastly, for a subset $s\subset\{1,2,\cdots,m\}$ we define $\mathcal{F}(s):=\{p\in\Delta\cap M\ |\ s\subset S(p)\}$ and $\hat{\mathcal{F}}(s):=\mathcal{F}(s)\cap\text{vert}(\Delta)$ with $\text{vert}(\Delta)$ the set of vertices of $\Delta$. We proceed as follows.

\vspace{3mm}
\setlength{\algomargin}{0em}
\begin{algorithm}[H]
    \begin{enumerate}[labelindent=0pt]
        \item Let $\mathfrak{F}_{n-1}=\{\{1\},\{2\},\cdots,\{m\}\}$. An element $s$ of this set represents a facet of the polytope with points $\mathcal{F}(s)$ and vertices $\hat{\mathcal{F}}(s)$.
        \item Recursively find the set of $i$-dimensional faces using
        $$\mathfrak{F}_{i}=\bigl\{f\cup g\ |\ f,g\in \mathfrak{F}_{i+1}\text{ and affdim}(\hat{\mathcal{F}}(f\cup g))=i\bigr\},$$
        where $\text{affdim}(\hat{\mathcal{F}}(f\cup g))$ is the dimension of the affine lattice spanned by the points $\hat{\mathcal{F}}(f\cup g)$.
        \item (optional) For computational efficiency, the computation of $\mathfrak{F}_{i}$ with $i\leq 2$ can be slightly simplified as follows.
        $$\mathfrak{F}_{i}=\bigl\{f\cup g\ |\ f,g\in \mathfrak{F}_{i+1}\text{ and }|\hat{\mathcal{F}}(f\cup g)|\geq i+1\bigr\}.$$
    \end{enumerate}
    \caption{Find the faces of $\Delta$.}
    \label{algo:faces}
\end{algorithm}
\vspace{3mm}

Representing a face as  a subset $f\subset\{1,2,\cdots,m\}$ allows for very efficient computation of its lattice points, filtering between interior or boundary lattice points, and finding the faces of the face. All of these computations are handled in \texttt{CYTools} by the \mintinline{python}{PolytopeFace} class.

This concludes our brief account of the main computations available for lattice polytopes. There are many more computations that are available in \texttt{CYTools}, which we omit here for brevity. The full list of functions available for the \mintinline{python}{Polytope} and \mintinline{python}{PolytopeFace} classes can be found in our documentation website \cite{cytools-website}.

\subsection{Triangulations}\label{sec:methods_triangulations}

Triangulations of polytopes are handled in \texttt{CYTools} by the \mintinline{python}{Triangulation} class, which contains a variety of functions to compute many key properties of triangulations, and also serves to interface with external software that performs other computations. The key external software packages used by this class are \texttt{TOPCOM} \cite{Rambau2002}, \texttt{Qhull} \cite{Barber96}, and the triangulations module \cite{cgal:dDTriangs} of the Computational Geometry Algorithms Library (\texttt{CGAL}) \cite{cgal:eb-20a}. We will describe the some of the roles these pieces of software play in the following sections.

\subsubsection{Triangulation algorithms}\label{sec:triangulation_algorithms}

As previously mentioned, \texttt{TOPCOM}, \texttt{Qhull}, or \texttt{CGAL} can be used to construct triangulations. \texttt{CGAL} is the default backend, but another backend can be specified by setting the \mintinline{python}{backend} keyword to the strings \mintinline{python}{"topcom"}, \mintinline{python}{"qhull"}, or \mintinline{python}{"cgal"}, respectively. When using \texttt{TOPCOM} as the backend, the triangulation that is obtained is the placing or pushing triangulation, which is a regular triangulation with specific properties (see \cite{DeLoera2010} for more details). When using \texttt{Qhull} or \texttt{CGAL} as the backend, the triangulation is obtained by lifting the points into an extra dimension by a set of heights $h_i$ and projecting the lower faces of the convex hull. Recall that regular triangulations are those that can be obtained in such a way, and thus these backends will generically produce regular triangulations. There are measure zero subsets of heights that produce polyhedral subdivisions that do not fully consist of simplices. Both \texttt{Qhull} and \texttt{CGAL} automatically subdivide further to obtain a triangulation, but only \texttt{CGAL} does so preserving regularity, and so it is preferred over \texttt{Qhull}. For this reason, \texttt{CGAL} defaults to the Delaunay triangulation, which uses heights $h_i=|\mathbf{p}_i|^2$, where $\mathbf{p}_i$ is the location of the $i$th lattice point.
\texttt{CGAL}  then subdivides further if necessary, whereas \texttt{Qhull} defaults to a nearby triangulation using heights $h_i=|\mathbf{p}_i|^2+\varepsilon_i$ with $\varepsilon_i$ chosen from a narrow Gaussian distribution.

As we showed in the example in \S\ref{sec:obtaining_triangulation}, the default triangulation that is obtained for reflexive polytopes is a fine, regular, star triangulation. We have already discussed why \mintinline{python}{triangulate} always returns a regular triangulation, so now let us discuss the other two properties. A placing or pushing triangulation is always fine, and a triangulation obtained from heights very close to the Delaunay heights will also be fine \cite{DeLoera2010}. The way we ensure that triangulations are star, when using \texttt{Qhull} or \texttt{CGAL}, is by decreasing the height of the origin until it is much lower than the rest.  When using \texttt{TOPCOM} we first obtain a non-star triangulation and turn it into a star triangulation by deleting internal lines and connecting all the points to the origin (see \cite{Braun:2014xka}), which is equivalent to decreasing the height of the origin when it is constructed in terms of heights.

\subsubsection{Sampling algorithms}\label{sss:sample}
Now we discuss algorithms to produce large sets of triangulations, starting with the simplest such procedure. As we discussed above, there is a special kind of fine, regular triangulation called the Delaunay triangulation, which can also be made star easily. This triangulation can be obtained by picking heights $h_i=|\mathbf{p}_i|^2$, and then subdividing further into simplices if necessary. We can thus use the Delaunay triangulation as a starting point to generate random triangulations, as follows.

\vspace{3mm}
\setlength{\algomargin}{0em}
\begin{algorithm}[H]
    \begin{enumerate}
        \item Take a Gaussian distribution with standard deviation $\sigma$. The parameter $\sigma$ will control how diverse the generated triangulations are, at the expense of how often they are not fine.
        \item Set an initial height vector $h_i=|\mathbf{p}_i|^2$ and decrease the height of the origin by a large number so that all resulting triangulations are star.
        \item Pick random elements $\varepsilon_i$ from the Gaussian distribution, and construct the triangulation given by the heights $h_i+\varepsilon_i$. If the triangulation is fine, store it; if not, discard it and repeat this step.
        \item Repeat step 3 until the desired number of triangulations is obtained.
    \end{enumerate}
    \caption{Fast (albeit unfair) generation of random triangulations}
    \label{algo:random_triangs_fast}
\end{algorithm}
\vspace{3mm}
This algorithm is very fast, and at large $h^{1,1}$ it can produce a practically unlimited set of triangulations.
However, the triangulations that it obtains are not fairly sampled among the full set of triangulations. Thus, even though it can be useful for some things such as machine learning, it should not be used to draw conclusions about distributions in the landscape. Performing a fair sampling of triangulations is a difficult issue and requires a more elaborate setup, which we described in a previous paper \cite{Demirtas:2020dbm}. Here we recall an algorithm that we found to produce a fair sample at a moderately high $h^{1,1}$, where we had access to the full list of triangulations for calibration purposes.

\vspace{3mm}
\setlength{\algomargin}{0em}
\begin{algorithm}[H]
    \begin{enumerate}
        \item Choose a height vector $\mathbf{h}=\mathbf{h}_0$ that generates a fine triangulation. This can be taken to be a triangulation near the Delaunay triangulation.
        \item Choose a random unit vector $\boldsymbol{\alpha} \in \mathbb{R}^n$.
        \item Find the largest $\varepsilon = \varepsilon_\text{max}$ such that
        $\mathbf{h}' = (\mathbf{h} + \epsilon \boldsymbol{\alpha}) / |\mathbf{h} + \epsilon \boldsymbol{\alpha}|$ generates a fine triangulation.
        \item Choose a random real number $\delta \in [0, \varepsilon_\text{max}]$ and send $\mathbf{h} \to (\mathbf{h} + \delta \boldsymbol{\alpha}) / |\mathbf{h} + \delta \boldsymbol{\alpha}|$.
        \item Repeat $N_\text{walk}$ times the steps 2-4.
        \item Compute the triangulation $\cT_{\mathbf{h}}$ generated by the heights $\mathbf{h}$.
        \item Perform $N_\text{flips}$ random bistellar flips starting from $\cT_{h}$, keeping the triangulation
        fine and regular at each step. Call the result
        $\cT_{\text{new}}$.
        \item Make $\cT_\text{new}$ into a star triangulation, and store it in the ensemble.
        \item Repeat steps 2-8 to accumulate the number of triangulations desired.
    \end{enumerate}
    \caption{Fair (albeit slow) generation of random triangulations \cite{Demirtas:2020dbm}}
    \label{algo:random_triangs_fair}
\end{algorithm}
\vspace{3mm}
This algorithm is fairly slow, as it requires some non-trivial steps such as checking regularity after each bistellar flip, and it works better at large $h^{1,1}$, as most of the directions in the region of FRSTs in the secondary fan are bounded. For a more in-depth discussion see \cite{Demirtas:2020dbm}, where we also obtained some distributions from the 4-dimensional reflexive polytopes with the largest Hodge numbers.

\subsubsection{Checking regularity}

Ensuring regularity of the triangulations is crucial in constructions of Calabi-Yau hypersurfaces, since, as we discuss in \S\ref{sec:methods_cones}, it is in correspondence with the variety being projective, and thus having a K\"ahler structure \cite{oda1991}.

The algorithm to determine whether a given triangulation is regular is most easily understood in terms of the secondary cone of the triangulation \cite{Gelfand1989,Gelfand1991,Gelfand1994}.
Remember that a triangulation of a point set is called regular if and only if it can be obtained by lifting the points into one extra dimension by a set of heights $h_i$, obtaining the convex hull of the resulting points, and projecting down the lower faces of the polyhedron. The subset of heights that give rise to a particular regular triangulation is a cone, called the secondary cone, since scaling all heights by a constant factor $\lambda \ge 0$ does not change the triangulation. A given triangulation is regular if and only if its secondary cone is full dimensional.

The secondary cone can be obtained using an algorithm due to Oda and Park \cite{oda1991}. If the triangulation is fine, one can instead use the much more efficient algorithm by Berglund, Katz and Klemm \cite{Berglund1995}, see also \cite{Altman:2014bfa}. Once the secondary cone is constructed, we check whether it is full dimensional in various ways. For low-dimensional cones, this can be done using \texttt{PPL}. However, for high-dimensional cones one must use numerical methods to explicitly find a point inside the cone. For this purpose, \texttt{CYTools} uses multiple approaches, including the \texttt{GLOP} solver from \texttt{OR-Tools} \cite{ortools}, and \texttt{MOSEK} \cite{mosek}.

\subsection{Calabi-Yau manifolds}

Finally, we describe the algorithms used by \texttt{CYTools} to analyze Calabi-Yau manifolds themselves.

\subsubsection{Intersection numbers}\label{sec:methods_intnums}

The intersection numbers are among the topological data most important for string compactifications, and \texttt{CYTools} can compute them many orders of magnitude faster than previously available software could. We will now describe the algorithm for computing the intersection numbers and comment on how the dramatic increase in efficiency was achieved. For concreteness, we will focus on computing the intersection numbers of a favorable Calabi-Yau threefold hypersurface.

The first step --- and the place where the computational challenge occurs --- is the computation of the intersection numbers $\kappa_{ijkl}^V$ of the ambient variety $V \supset X$. Once $\kappa_{ijkl}^V$ are known, the intersection numbers of the Calabi-Yau hypersurface are simply,
\begin{equation}
    \kappa_{ijk}^X = \sum_{l=1}^{h^{1,1}+4} \kappa_{ijkl}^V\,.
\end{equation}
The intersection numbers of distinct divisors, i.e.~$\kappa_{ijkl}^V, i\neq j \neq k \neq l$ are determined by the volumes of full dimensional cones in the toric fan. Specifically, consider the toric cone generated by the rays $(v_i, v_j, v_k, v_l)$, corresponding to toric divisors $(D^i, D^j, D^k, D^l)$. The properly normalized volume of this cone is given by the determinant of the $4\times 4$ matrix
\begin{equation}\label{eqn:mat1}
M_{ijkl}=
  \begin{pmatrix}
  v_i^1 & v_i^2 & v_i^3 & v_i^4 \\
  v_j^1 & v_j^2 & v_j^3 & v_j^4 \\
  v_k^1 & v_k^2 & v_k^3 & v_k^4 \\
  v_l^1 & v_l^2 & v_l^3 & v_l^4
  \end{pmatrix}.
 \end{equation}
The distinct intersection numbers are then
\begin{equation}
    \kappa_{ijkl}^V = \frac{1}{\det(M_{ijkl})}, \quad i\neq j \neq k \neq l.
\end{equation}
The last step is to calculate the self-intersection numbers. Remember that there are 4 linear equivalence relations between the toric divisors,
\begin{equation}
    L^a_i \hat{D^i} = 0,
\end{equation}
with $a=1 \dots 4$. We can multiply this linear system by $\hat{D}^j \cap \hat{D}^k \cap \hat{D}^l$ to obtain
\begin{equation}
    L^a_i \, \hat{D^i} \cap \hat{D}^j \cap \hat{D}^k \cap \hat{D}^l = L^a_i \kappa_{ijkl}^V = 0.
\end{equation}
One can then iterate over all possible choices of the triplet $\{i,j,k\}$ to construct a large linear system, involving all intersection numbers $\kappa_{ijkl}^V$. This system is over-determined and has a unique solution.

One might worry that such a linear system would be unwieldy, as the number of equations is $4(h^{1,1}+4)^3$. There are two crucial factors that render this system solvable, even when $h^{1,1} \sim \mathcal{O}(1000)$. First, if $v_i$ and $v_j$ are not on the same facet, the corresponding toric divisors cannot intersect. So, the linear equation from any triplet $\{i,j,k\}$ is trivial unless the corresponding rays are on the same facet. This reduces the size of the linear system considerably. Second, the linear system is extremely sparse and can be solved via a Cholesky decomposition. This enables the use of sparse linear algebra tools, such as \cite{10.1145/1391989.1391995}.

\subsection{Cones}\label{sec:methods_cones}

Let us conclude by discussing how we perform some computations involving cones. The basic operation of switching between ray and hyperplane representations is done with the Parma Polyhedra Library (\texttt{PPL}) \cite{BagnaraHZ08SCP}.
This software provides excellent performance, but because of the inherent exponential growth of complexity with the dimension of the cone, the change of representation can only be done for cones of relatively low dimensions, or for particularly simple high-dimensional cones. Nevertheless, one can avoid a change of representation for most computations of interest. We have developed the necessary tools to analyze high-dimensional cones, to the point where even the highest-dimensional Mori and K\"ahler cones, with 491 dimensions, can be easily studied.

\subsubsection{Finding Mori and K\"ahler cones}

The Mori and K\"ahler cones of a toric variety are closely related to the secondary cone of the triangulation \cite{oda1991}. This can be understood from the fact that the K\"ahler cone of the toric variety can be thought of a projection of the secondary cone that removes the linear subspaces. As such, a regular, star triangulation is in correspondence with a projective variety, with a non-empty K\"ahler cone. With this knowledge, we can find both the Mori and K\"ahler cones by using the algorithm from \cite{Berglund1995} that we mentioned in \S\ref{sec:methods_triangulations}.

\subsubsection{Finding extremal rays}

Suppose that we start with a cone generated by a set $\mathcal{S}$ of rays. The extremal rays are the minimal set of generators of the cone. In particular, an extremal ray cannot be written as a non-negative linear combination of the remaining rays. The question of whether a ray is extremal can be difficult to answer when using exact rational arithmetic. However, it very efficient to use a numerical approach, in particular via a non-negative least squares minimization. Let $\mathbf{b}$ be the ray of interest, and $\mathbf{A}$ be the matrix whose columns are the remaining generating rays. We minimize $|Ax-b|$ subject to $x_i\geq0$. If the result is equal to zero (up to some tolerance) then the ray is not extremal, and otherwise it is extremal. For the minimization we use the Non-Negative Least Squares (NNLS) solver in SciPy, which is taken from \cite{doi:10.1137/1.9781611971217}. This procedure works well, even for high-dimensional cones. Furthermore, the procedure is parallelized when the cone is pointed. For non-pointed cones, one must be careful with the rays generating linear subspaces, so the process must be single-threaded. Moreover, for cones that are very high-dimensional, and additionally are extremely wide or narrow, it can sometimes misidentify rays or fail. Nevertheless, it is still far better than other approaches known to us.

\subsubsection{Finding the tip of a stretched cone}
Let $\mathcal{C}$ be an $n$-dimensional polyhedral cone defined by $m$ linear inequivalences
\begin{equation}
    \mathcal{C} := \Bigl\{\vec{x} \in \mathbb{R}^n \,\Bigl|\, \textbf{M} \cdot \vec{x} \geq 0 \Bigr\},
\end{equation}
where $\textbf{M}$ is an $n \times m$ matrix. We define the $c$-stretched cone $\widetilde{\mathcal{C}}[c]$ to be
\begin{equation}
    \widetilde{\mathcal{C}}[c] := \Bigl\{\vec{x} \in \mathbb{R}^n \,\Bigl|\, \textbf{M} \cdot \vec{x} \geq c \Bigr\}.
\end{equation}
Note that this is a slight abuse of language, as $\widetilde{\mathcal{C}}[c]$ is not always a cone. We call the point $\vec{x} \in \widetilde{\mathcal{C}}[c]$ that is  closest to the origin the tip of $\widetilde{\mathcal{C}}[c]$. Finding this point requires solving the problem
\begin{align}
    \text{minimize} \qquad &|\vec{x}|^2 \\
    \text{subject to} \qquad &\textbf{M} \cdot \vec{x} \geq c.
\end{align}
This is an instance of a quadratic programming problem, and can be solved efficiently, even when $n \sim \mathcal{O}(1000)$. \texttt{CYTools} provides access to multiple backends for solving such problems, including \texttt{MOSEK} \cite{mosek} and \texttt{OSQP} \cite{osqp}.

\section{Outlook}\label{sec:outlook}

\texttt{CYTools} provides fast, automated computation of the topological data of Calabi-Yau hypersurfaces and their parent toric varieties.
It has powerful tools for triangulating polytopes and for the cone and lattice problems that are commonplace in Calabi-Yau compactifications.  In many individual problems in this area,
\texttt{CYTools} is substantially faster than prior software packages.  For the complete, end-to-end analysis from a typical (or, especially, a difficult) polytope in the Kreuzer-Skarke list to the classical data of a 4-dimensional effective theory, \texttt{CYTools} represents an improvement in speed by many orders of magnitude.
The numbers of configurations arising in many steps are exponential in the relevant Hodge numbers, and general-purpose computational geometry algorithms are effective only for Hodge numbers $\lesssim 10$.  \texttt{CYTools} can handle the largest polytope in the Kreuzer-Skarke list, with $h^{1,1}=491$, in seconds.

\texttt{CYTools} integrates into a single platform many capabilities used in the analysis of string compactifications.
We hope that \texttt{CYTools} will contribute to accelerating progress in the field in several ways.  First, it can remove the many small barriers presented by the need to code up elementary operations.  Second, when multiple researchers (or research groups) studying related problems use \texttt{CYTools} format, their codes are interoperable without translation, aiding reproducibility.  For example, the code shown in \S\ref{sec:exscan} allows one to reproduce the key results of \cite{Demirtas:2018akl}.  Finally, \texttt{CYTools} is fast enough to make analysis of the entire Kreuzer-Skarke list feasible for the first time, opening the door on a vast landscape.

The computational capabilities afforded by \texttt{CYTools} have already been applied to a range of problems in string compactifications: see e.g.~\cite{Demirtas:2020dbm,Demirtas:2020ffz,Mehta:2021pwf,Cicoli:2021dhg,Demirtas:2021nlu,Kim:2021lwo,Demirtas:2021gsq,Grana:2022dfw,Gendler:2022qof,Crino:2022zjk,Tsagkaris:2022apo}.
Let us comment on a few examples in order to illustrate the sorts of problems that are accessible.
\texttt{CYTools} was used in  \cite{Cicoli:2021dhg}
to study del Pezzo divisors in an ensemble of  Calabi-Yau threefolds with $h^{1,1} \le 40$, and was used in
\cite{Crino:2022zjk} to construct orientifolds for all favorable Calabi-Yau threefolds with $h^{1,1} \le 7$.
The statistics of axion couplings were obtained in large ensembles reaching up to $h^{1,1} = 491$ in \cite{Demirtas:2018akl,Mehta:2021pwf,Demirtas:2021gsq}, with applications to black hole superradiance in \cite{Mehta:2021pwf} and to the strong CP problem in \cite{Demirtas:2021gsq}.
Moduli stabilization and the contributions of quantized fluxes to the D3-brane tadpole were studied in \cite{Grana:2022dfw} and \cite{Tsagkaris:2022apo},
and \texttt{CYTools} facilitated the study of cases with large Hodge numbers.
The study of perturbatively flat vacua in \cite{Demirtas:2019sip,Demirtas:2020ffz} and especially of
supersymmetric AdS vacua in \cite{Demirtas:2021nlu} relied heavily on \texttt{CYTools}.  In sum, \texttt{CYTools} facilitates automated searches through the landscape of Calabi-Yau threefold hypersurfaces
and of flux vacua therein,
allowing for exploration and enumerative experiment.

There are several obvious directions for future development of \texttt{CYTools}. One is to encompass a broader class of Calabi-Yau threefolds, such as complete intersections in toric varieties.
Another is to construct orientifolds automatically, and, where applicable, to relate them to compactifications of F-theory on Calabi-Yau fourfolds.  Although \texttt{CYTools} can analyze fourfolds, it would be worthwhile to develop richer capabilities in this area.
Another important step will be to compute the topological data that governs nonperturbative quantum effects in string compactifications, such as the Gopakumar-Vafa invariants of curves, and the zero modes and Pfaffians of Euclidean D-branes.  Finally, one could extend \texttt{CYTools} to compute the zeta-functions of Calabi-Yau threefolds defined over the integers.

We hope to incorporate progress in some of these directions in a future version.  Moreover, we invite collaboration from members of the community who would like to integrate a wider range of computational capabilities into \texttt{CYTools}.

\section*{Acknowledgements}

We are grateful to Nate MacFadden, Jakob Moritz, and Andreas Schachner for finding many bugs in
\texttt{CYTools}.
We thank Naomi Gendler, Manki Kim, Dnyanesh Kulkarni, Jakob Moritz, Richard Nally, and Andreas Schachner for comments on a draft of this work, and we are grateful to
Gauri Batra, Arjun Chaturvedi, Geoffrey Fatin,  Jim Halverson, Nate MacFadden, Viraf Mehta,   Brent Nelson, Murali Saravanan, and Sam Weiss for discussions of related topics.
Furthermore, we thank Andreas Braun and Sven Krippendorf for providing inspiration to pursue this work.
We are grateful to Andreas Braun, Cody Long, Mike Stillman, and Ben Sung for their essential contributions to predecessors of this work, parts of which appeared in \cite{Long:2014fba,Braun:2017nhi,Demirtas:2018akl}.
We are particularly indebted to Mike Stillman for explaining to us the algorithm for computing intersection numbers that is now used in \texttt{CYTools}, and for patiently teaching us countless pieces of mathematics needed for this work.
Finally, we thank Manki Kim and Jakob Moritz for their collaboration in developing new capabilities to compute Gopakumar-Vafa invariants in \texttt{CYTools} \cite{computational-mirror-symmetry}. The research of M.D.~was supported in part by the National Science Foundation under Cooperative Agreement PHY-2019786 (The NSF AI Institute for Artificial Intelligence and Fundamental Interactions). The research of L.M.~and A.R.-T.~was supported in part by the National Science Foundation through the grant PHY-2014071.

\appendix
\addtocontents{toc}{\protect\setcounter{tocdepth}{1}}

\section{List of Terms}\label{appx:definitions}

\subsection{Polytopes}
\begin{itemize}
    \item A \emph{lattice} $M$ of rank $n$ is a finitely generated free abelian group of rank $n$, which we always think of as $\mathbb{Z}^n$. We denote the associated real vector space as $M_{\mathbb{R}}:=M\otimes_{\mathbb{Z}}\mathbb{R}$.
    \item The \emph{dual lattice} $N$ of a lattice $M$ is the space of homomorphisms from $M$ to $\mathbb{Z}$, $N:=\mathrm{Hom}(M,\mathbb{Z})$, which is also isomorphic to $\mathbb{Z}^n$.
    \item A \emph{polytope} $\Delta\subset M_{\mathbb{R}}$ is the convex hull of a finite set of points $S\subset M_{\mathbb{R}}$, i.e. $\Delta=\mathrm{Conv}(S)$.
    \item The \emph{dimension} $d$ of a polytope $\Delta$ is the dimension of the smallest affine subspace of $M_{\mathbb{R}}$ containing $\Delta$. If it has the same dimension as its ambient space then it is called \emph{full-dimensional} or \emph{solid}.
    \item A \emph{lattice polytope} $\Delta$ is a polytope that is the convex hull of a finite set of lattice points $S\subset M$.
    \item The \emph{dual polytope} $\Delta^\circ$ of a full-dimensional polytope $\Delta$ that contains the origin is the convex hull $\Delta^\circ := \mathrm{Conv}(\{y\in N_{\mathbb{R}}\ |\ y\cdot x\geq -1 \text{ for all } x\in\Delta\})$.
    \item A polytope $\Delta$ is called \emph{reflexive} if both $\Delta$ and $\Delta^\circ$ are lattice polytopes.
    \item A subset $F\subseteq\Delta$ is called a \emph{face} of $\Delta$ if there are $u\in N_{\mathbb{R}}\setminus\{0\}$ and $b\in\mathbb{R}$ such that $F=\{m\in M_{\mathbb{R}}\ |\ m\cdot u=b\}$ and $\Delta\subset \{m\in M_{\mathbb{R}}\ |\ m\cdot u\geq b\}$. The empty set and $\Delta$ itself are faces of $\Delta$. A face of codimension $1$ in the polytope is called a \emph{facet}, a face of dimension $1$ is called an \emph{edge}, and a face of dimension $0$ is called a \emph{vertex}.
    \item  A lattice point of $\Delta$ is called a \emph{boundary point} if it is contained in a facet, and is otherwise called an \emph{interior point}.  Note that when $\Delta$ is reflexive, the only interior point is the origin.
    \item A polytope of dimension $n$ is called a \emph{simplex} or an $n$-\emph{simplex} if it has exactly $n+1$ vertices, and is called \emph{simplicial} if every facet is an $n-1$-simplex.
\end{itemize}

\subsection{Triangulations}
\begin{itemize}
    \item A \emph{triangulation} $\mathcal{T}$ of a lattice polytope $\Delta^\circ$ is a collection of simplices in $N_{\mathbb{R}}$ satisfying:
    \begin{itemize}
        \item Each simplex has the same dimension as $\Delta^\circ$ and has vertices in $\Delta^\circ\cap N$.
        \item The intersection of any two simplices in $\mathcal{T}$ is a face of each.
        \item The union of all simplices in $\mathcal{T}$ is $\Delta^\circ$.
    \end{itemize}
    \item A triangulation of a point set is called \emph{fine} if every point is the vertex of some simplex in the triangulation.
    When the point set is a 4-dimensional reflexive polytope $\Delta^\circ$, we call a triangulation $\mathscr{T}$ fine even if $\mathscr{T}$ omits some or all
     of the points that are strictly interior to facets of $\Delta^\circ$, since the inclusion of these points is not required to obtain smooth Calabi-Yau hypersurfaces \cite{Batyrev:1994hm}.
    \item A triangulation is called \emph{star} if the origin is a vertex of all the simplices of the triangulation. Such a triangulation can be interpreted as defining a refinement of the face fan of the polytope.
    \item A triangulation is called \emph{regular} if it can be obtained as the projection of the lower facets of the convex hull $\{(\vec{p}_i,h_i)\in N_{\mathbb{R}}\times\mathbb{R}\ |\ \vec{p}_i\in N_{\mathbb{R}}\cap\Delta^\circ\}$ for some set of \emph{heights} $h_i \in \mathbb{R}$.
\end{itemize}

\subsection{Cones}
Let $X$ be a projective algebraic variety of complex dimension $n$.
\begin{itemize}
    \item An effective divisor on $X$ is
    a finite formal sum, with nonnegative coefficients, of irreducible codimension-1 subvarieties.
    The \emph{effective cone}, or cone of effective divisors, is the convex cone in $H_{2n-2}(X,\mathbb{R})$ generated by the classes of effective divisors on $X$.
    \item The \emph{Mori cone} of $X$ is the convex cone in $H_2(X,\mathbb{R})$ generated by holomorphic curves.  That is, it is the cone of effective subvarieties of complex dimension one.
    \item The \emph{K\"ahler cone} is the cone of cohomology classes of K\"ahler forms in
    $H^{1,1}(X,\mathbb{R})$.  The closure of the K\"ahler cone is the dual of the Mori cone.
\end{itemize}

\bibliographystyle{JHEP}
\bibliography{refs}

\end{document}